\documentclass[aps,prb,10pt,twocolumn,amsmath,amssymb,showpacs,
longbibliography,superscriptaddress]{revtex4-1}

\usepackage{amsmath,amssymb,amsfonts} 	
\usepackage{graphicx}
\usepackage{hhline}
\usepackage[latin1]{inputenc}
\usepackage{subfigure}
\usepackage{hyperref}
\usepackage{bm}
\usepackage{bbm}
\usepackage{nicefrac}

\newcommand{\eq}[1]{Eq.~(\ref{eq:#1})}
\newcommand{\eqs}[2]{Eqs.~(\ref{eq:#1}) and~(\ref{eq:#2})}
\newcommand{\equ}[1]{Equation~(\ref{eq:#1})}

\newcommand{\bE}{{\bm E}}
\newcommand{\bB}{{\bm B}}
\newcommand{\bP}{{\bm P}}
\newcommand{\bM}{{\bm M}}
\newcommand{\bk}{{\bm k}}
\newcommand{\br}{{\bm r}}
\newcommand{\bn}{{\bm n}}
\newcommand{\ba}{{\bm a}}
\newcommand{\trev}{{\cal T}}
\newcommand{\parity}{{\cal P}}

\def\nn{\nonumber\\}
\def\beq{\begin{equation}}
\def\eeq{\end{equation}}
\def\bea{\begin{eqnarray}}
\def\eea{\end{eqnarray}}
\def\ket#1{\vert#1\rangle}
\def\bra#1{\langle#1\vert}
\def\ip#1#2{\langle#1\vert#2\rangle}
\def\me#1#2#3{\langle#1\vert#2\vert#3\rangle}
\def\wt#1{\widetilde{#1}}
\def\kpar{{\boldsymbol\kappa}}
\def\tzero{\theta_0}
\def\ttwo{\theta_2}
\def\tthree{\theta_3}
\def\Z2{\mathbbm{Z}_2}
\def\la{\langle\kern-2.0pt\langle}
\def\ra{\rangle\kern-2.0pt\rangle}
\def\hwf{\mathfrak{h}}
\def\tint{t_3}
\def\tintl{t_{3,p}}
\def\gbar{\overline{\Gamma}}
\def\kbar{\overline{\rm K}}
\def\kbar{\overline{\rm K}}
\def\mbar{\overline{\rm M}}
\def\gbar{\overline{\Gamma}}

\def\cycle{\phi} 

\begin{document}

\title{Surface theorem for the Chern-Simons axion coupling}

\author{Thomas Olsen} \email{tolsen@fysik.dtu.dk} \affiliation{Center
  for Atomic-Scale Materials Design, Department of Physics, Technical
  University of Denmark, 2820 Kgs. Lyngby Denmark} \affiliation{Center
  for Nanostructured Graphene (CNG), Department of Nanotechnology,
  Technical University of Denmark, 2820 Kgs. Lyngby Denmark}

\author{Maryam Taherinejad} \affiliation{Materials Theory, ETH
  Z{\"u}rich, Wolfgang-Pauli-Strasse 27, CH-8093 Z{\"u}rich,
  Switzerland}

\author{David Vanderbilt} \affiliation{Department of Physics
  Astronomy, Rutgers University, Piscataway, New Jersey 08854-8019,
  USA}

\author{Ivo Souza} \affiliation{Centro de F{\'i}sica de Materiales,
  Universidad del Pa{\'i}s Vasco, 20018 San Sebasti{\'a}n, Spain}
\affiliation{Ikerbasque Foundation, 48013 Bilbao, Spain}

\date{\today}
\begin{abstract}
  The Chern-Simons axion coupling of a bulk insulator is only defined
  modulo a quantum of $e^2/h$.  The quantized part of the coupling is
  uniquely defined for a bounded insulating sample, but it depends on
  the specific surface termination. Working in a slab geometry and
  representing the valence bands in terms of hybrid Wannier functions,
  we show how to determine that quantized part from the excess Chern
  number of the hybrid Wannier sheets located near the surface of the
  slab. The procedure is illustrated for a tight-binding model
  consisting of coupled quantum anomalous Hall layers.  By slowly
  modulating the model parameters, it is possible to transfer one unit
  of Chern number from the bottom to the top surface over the course
  of a cyclic evolution of the bulk Hamiltonian. When the evolution of
  the surface Hamiltonian is also cyclic, the Chern pumping is
  obstructed by chiral touchings between valence and conduction
  surface bands.
\end{abstract}
\maketitle

\section{Introduction}

The axion field was originally introduced as a strategy for resolving
the non-violation of time reversal ($\trev$) and spatial inversion
($\parity$) symmetry in quantum chromodynamics
(QCD).\cite{peccei-prl77} Specifically, the gauge symmetry of QCD
allows a term in the Lagrangian that is a non-Abelian version of
\beq
\label{eq:L-theta}
\mathcal{L}_\theta=\frac{ e^2}{h}\frac{\theta}{2\pi}
\bE\cdot\bB,
\eeq
where $\theta$ is a fixed dimensionless parameter, and $\bE$ and $\bB$
are the electric and magnetic fields. Since $\bB$ and $\bE$ are odd
under $\trev$ and $\parity$ respectively, these symmetries are
individually broken by $\mathcal{L}_\theta$. The most striking
consequence of such a term is the prediction of a finite electric
dipole moment for the neutron, but recent experimental bounds on this
quantity\cite{pendlebury-prd15} restrict $|\theta|$ to be less than
about $10^{-9}$.  Such fine tuning is regarded as unnatural, and in
order to circumvent it Peccei and Quinn showed that promoting $\theta$
to a dynamical field leads to a vanishing vacuum expectation value for
the field ($\theta=\langle\hat\theta\rangle=0$), elegantly restoring
$\trev$ and $\parity$ symmetry in QCD.\cite{peccei-prl77} Excitations
of the field naturally give rise to a new massive particle known as
the axion.\cite{weinberg-prl78, wilczek-prl78} The axion has never
been observed, but is presently regarded as an important candidate for
dark matter.\cite{asztalos-prl10}

The classical field theories resulting from the inclusion of the
$\mathcal{L}_\theta$ term have several interesting properties and have
been investigated in a number of different
contexts.\cite{wilczek-prl87} A most remarkable property is that the
theory is invariant under $\theta\rightarrow\theta+2\pi$, so that
$\theta$ is best viewed as a phase. This is related to the fact that
$\bE\cdot\bB=\varepsilon_{abcd}F^{ab}F^{cd}/8$ is proportional to the
second Chern class of the gauge field. Hence, the integral of
$\mathcal{L}_\theta/\theta$ over a closed manifold yields an integer
that represents the winding number of the field
configuration.\cite{witten-rmp16} In addition, the evaluation of any
correlation function involves a path integral of
$\exp[i\int dx^4 \mathcal{L}_\theta(x)]$, which becomes
$e^{i n\theta}=e^{i n(\theta+2\pi)}$ for any field configuration that
vanishes at infinity.

Another interesting interpretation of $\theta$ relates to the
nontrivial vacuum structure allowed by non-Abelian gauge
theories. Different vacua are then classified by the winding number
$n$ of classical field configurations, and the vacuum state is taken
as $|\theta\rangle=\sum_ne^{i\theta
  n}|n\rangle$.\cite{gabadadze-ijmp87} In this context $\theta$ is
referred to as the vacuum angle, and can be regarded as the ``Bloch
momentum'' associated with a particular state on the ``lattice'' of
topological distinct field configurations.

In a condensed-matter setting, the analogue of a vacuum state is the
electronic ground state of an insulator. In that case $\theta$ is a
material property determined by the electronic structure.  If we allow
the presence of $\mathcal{L}_\theta$ in the Lagrangian, Maxwell's
equations acquire additional $\theta$-dependent terms. Since
$\mathcal{L}_\theta$ with constant $\theta$ can be rewritten as a
total derivative, only regions of changing $\theta$ can give rise to
physically observable effects, as long as the gauge fields are treated
classically. If one now considers an insulating boundary separating
two materials characterized by different $\theta$ values, it is
straightforward to show that the interface will support an in-plane
current given by\cite{wilczek-ps16, qi2008}
\beq\label{eq:j_theta} 
j_i=\widetilde j_i + \frac{e^2}{h}\frac{\Delta\theta}{2\pi}
\varepsilon_{ij}E^j.
\eeq 
The first term is the usual current appearing in
Maxwell's equations, and the second is an additional ``axial'' current
that arises due to the change $\Delta\theta$ 
across the boundary. We now see that changing
$\Delta\theta\rightarrow\Delta\theta+2\pi$ amounts to adding a quantum
of anomalous Hall conductivity (AHC) to the interface. Conversely, if 
the AHC at an insulating interface is known, it becomes possible 
to assign a definite value to $\Delta\theta$.

The transformation properties of $\bE$ and $\bB$ in \eq{L-theta}
indicate that $\theta$ is odd under $\trev$ and $\parity$. Due to the
$2\pi$ ambiguity, those symmetries allow for $\theta=\pi$ as well as
$\theta=0$, leading to a $\Z2$ topological classification. It is well
known that when $\trev$~symmetry is present the value $\theta=\pi$
describes strong topological insulators, while topologically trivial
insulators have $\theta=0$.\cite{qi2008} In the case that $\parity$ is
conserved but not $\trev$, the term ``axion insulator'' has sometimes
been used\cite{wan-prb11,balents-p11} to describe the topological
phase with $\theta=\pi$. More generally, any magnetic point group that
contains a proper rotation composed with $\trev$, or an improper
rotation not composed with $\trev$, supports a $\Z2$ classification
with $\theta$ constrained to be~$0$ or~$\pi$.\cite{fang-prb12}

The existence or nature of topologically-protected boundary states
depends on the symmetry protecting $\theta=\pi$.  For example, when
$\trev$ remains unbroken at the surface of a strong topological
insulator, that surface is guaranteed to harbor an odd number of
metallic Dirac cones, with a half-quantized surface AHC that exactly
cancels the axial current.\cite{coh-prb11} In contrast, the surface of
an axion insulator automatically breaks $\parity$, so that surface
states are not protected. In the case of mirror symmetry, metallic
states are only protected on mirror-preserving surfaces, while
surfaces that break mirror symmetry can be insulating and display a
half-quantized AHC.

The term $\mathcal{L}_\theta$ given by \eq{L-theta} is not present
\textit{a priori} in the action of condensed-matter systems, where it
should be regarded as an effective term that sometimes provides a
useful description. Indeed, such a term appears when the electrons are
integrated out of a generic solid state partition function in four
dimensions, followed by dimensional reduction.\cite{qi2008} That
procedure leads to the expression
\begin{align}\label{eq:theta_3d}
\theta_{\rm CS}=-\frac{1}{4\pi}\int d\bk \, \epsilon_{ijl}
\text{Tr}
\left[
  A^i_\bk\partial_{k_j}A^l_\bk-i\frac{2}{3}A^i_\bk A^j_\bk A^l_\bk
\right]
\end{align}
for the axion coupling strength. Here 
\beq
\label{eq:connection-bloch}
A^i_{\bk nm}=i\ip{u_{\bk n}}{\partial_{k_i}u_{\bk m}}
\eeq
is the Berry connection matrix in Cartesian direction $i$, where
$\ket{u_{\bk n}}$ is the cell-periodic part of the Bloch function
$\ket{\psi_{\bk n}}$ of the $n$th occupied band,
$\langle\ldots\rangle$ denotes an integration over one crystal cell,
and the trace is over occupied bands. The integral in $d\bk$ is over
the Brillouin zone (BZ), and the integrand is known as the
Chern-Simons (CS) 3-form.

The form of \eq{L-theta} suggests a close relation between $\theta$
and the linear magnetoelectic response tensor defined as $\alpha_{ij}
=\left(\partial P_i/\partial B_j\right)_{\bE} =\left(\partial
  M_j/\partial E_i\right)_{\bB}$, where $\bP$ and $\bM$ are the
macroscopic polarization and magnetization respectively. With
$\widetilde{\alpha}$ the traceless part of $\alpha$, the relation
reads
\begin{subequations}
\begin{align}
\alpha_{ij}&=\widetilde{\alpha}_{ij}+\overline{\alpha}\delta_{ij}\\
\overline{\alpha}&=\frac{e^2}{h}\frac{\theta}{2\pi}.
\end{align}
\end{subequations}
This defines $\theta$ in terms of the trace piece $\overline{\alpha}$,
also known as the ``axion magnetoelectric coupling.''

The magnetoelectric response of a solid can be decomposed into spin
and orbital contributions on one hand, and frozen-ion and
lattice-mediated contributions on the other.  The frozen-ion orbital
part of the response was calculated for generic band insulators in
Refs.~\onlinecite{malashevich-njp10,essin2010}. It was found that the
trace piece takes the form $\overline{\alpha}_{\rm orb} =(e^2/2\pi
h)\left(\theta_{\rm CS}+\theta_{\rm cg}\right)$ where, in addition to
$\theta_{\rm CS}$ given by \eq{theta_3d}, there is a cross-gap term
$\theta_{\rm cg}$ that couples occupied and empty bands.  Among all of
the above terms, $\theta_{\rm CS}$ is the only term with a $2\pi$
ambiguity. It is thus sufficient, for the purpose of establishing the
surface theorem, to focus on the Chern-Simons axion (CSA) coupling
\beq 
\label{eq:alpha-cs}
\overline{\alpha}_{\rm CS}= \frac{e^2}{h}\frac{\theta_{\rm CS}}{2\pi}.
\eeq

Typical ground state properties of a band insulator (e.g., the charge
density and total energy) are invariant under any unitary
``gauge transformation''
\beq
\label{eq:gauge-transformation}
\ket{u_{\bk n}}\rightarrow \sum_m\ket{u_{\bk m}}U_{\bk mn}
\eeq
among the occupied Bloch orbitals. The $2\pi$ ambiguity in
$\theta_{\rm CS}$ comes about because \eq{theta_3d} changes by integer
multiples of $2\pi$ under certain gauge transformations.  The CSA
coupling will then change by an integer multiple of the quantum of
conductance. This is another manifestation of \eq{j_theta}: changing
$\theta_{\rm CS}$ by a multiple of $2\pi$ amounts to adding quantum
anomalous Hall layers at the surface, without modifying the
bulk.\cite{essin-prl09}

Another property of band insulators that behaves in this way is the
bulk polarization $\bP$, whose electronic part can be expressed as a
Berry phase.\cite{king-smith-prb93} The Berry phase may also change by
an integer multiple of $2\pi$ under unitary transformations, and the
quantum of ambiguity in the bulk definition of $\bP\cdot\hat{\bn}$ can
be resolved by taking into account quantized contributions of
$e/A_{\rm cell}$ to the surface charge density associated with an
insulating surface with orientation $\hat\bn$.\cite{vanderbilt-prb93}

In fact, $\overline{\alpha}_{\rm CS}$ and $\bP$ are related in more
than one respect. First of all, $\overline{\alpha}_{\rm CS}$ is in a
sense the natural three-dimensional generalization of the polarization
in one-dimensional systems, and is sometimes referred to as the
``magnetoelectric polarization.''\cite{qi2008} For any closed manifold
of odd dimension $d$, there exists a Chern-Simons $d$-form, which is a
functional of the $U(N)$ connection $A^i_{\bk mn}$. The 1-form is
simply the trace of the connection, and its BZ integral (the Berry
phase) is proportional to the polarization. Similarly, the BZ integral
of the Chern-Simons 3-form yields $\theta_{\rm CS}$, which is
proportional to the CSA coupling. Second, the $e^2/h$ ambiguity in
$\overline{\alpha}_{\rm CS}$ follows from the $e/A_{\rm cell}$
ambiguity in $\bP\cdot\hat\bn$.  \cite{essin2010} This is perhaps
surprising, since changes in polarization are well defined whenever
different states of polarization can be connected by an adiabatic
path.  However, as pointed out in Ref.~\onlinecite{essin2010}, if one
tries to define the CSA coupling as a response of $P_i$ to a smoothly
increasing $B_i$, a problem arises in that the smallest value of
$\Delta B_i$ compatible with the lattice periodicity corresponds to a
quantum of magnetic flux through the unit cell. The best one can do is
to define $\overline{\alpha}_{\rm CS}=\Delta P_i/\Delta B_i$ for a
single flux quantum, i.e., $\Delta B_i=h/eA_{\rm cell}$. Then
the gauge ambiguity of $e/A_{\rm cell}$ in $P_i$ implies a gauge
ambiguity of $e^2/h$ in $\overline{\alpha}_{\rm CS}$.

The surface theorem for polarization was derived in
Ref.~\onlinecite{vanderbilt-prb93}.  It states that the macroscopic
charge per surface unit cell at an insulating surface of a crystalline
insulator has, in units of $e$, a noninteger part that only depends on
the bulk polarization, and an integer part that is fixed by the
surface termination.

In this work, we demonstrate a similar surface theorem for the CSA
coupling of \eq{alpha-cs}. We take a bulk insulator with a given
$\theta_{\rm CS}$ (mod~$2\pi$), consider a specific insulating surface
termination in a slab geometry, and show how to use the knowledge
about the surface Hamiltonian to determine the CSA coupling exactly,
not just up to a quantum of $e^2/h$.

An important step towards that result was taken in
Ref.~\onlinecite{taherinejad-prl15}, where \eq{theta_3d} was recast in
terms of hybrid Wannier functions (HWFs), and then used to study a
``quantum CSA pump:'' a cyclic evolution of a bulk crystal in which
$\theta_{\rm CS}$ evolves continuously from some initial value
$\theta_{\rm CS}^i$ to reach $\theta_{\rm CS}^i+2\pi$ at the end of
the cycle. In the HWF representation, the pumping process is
transparent: the ``Wannier sheets'' carry quantized amounts of
Berry-curvature flux, and a $2\pi$ quantum gets transferred across
each unit cell via a sequence of sheet-touching events. These ideas
play a central role in the present work, where the HWF representation
will be used extensively.

The manuscript is organized as follows. In Sec.~\ref{sec:defs} we
establish the basic definitions and conventions used in this work, and
in Sec.~\ref{sec:hwf} we review the expression of
Ref.~\onlinecite{taherinejad-prl15} for $\theta_{\rm CS}$ in terms of
the bulk HWFs.  The goal of the ensuing subsections is to evaluate the
CSA coupling of a thick slab in the HWF basis.  The end result is the
surface theorem of \eq{surf-thm}, whereby the noninteger part of
$\theta_{\rm slab}/2\pi$ is given by the bulk expression in terms of
the HWFs in a unit cell deep inside the slab, and the integer part by
the net Chern number of the excess Wannier sheets near the surface. In
order to establish this result, we start in Sec.~\ref{sec:crystallite}
from an expression for the CSA coupling of a finite crystallite; then
in Sec.~\ref{sec:slab} we derive from it the corresponding expression
for a slab; finally in Sec.~\ref{sec:surf-thm} we consider its
limiting form for a thick slab. At every step all surface
contributions are carefully accounted for, leading to \eq{surf-thm}
for insulating slabs with vanishing Chern number. To validate the
derivation and illustrate the surface theorem, in
Sec.~\ref{sec:haldane} we study numerically a layered tight-binding
model that realizes a quantum CSA pump. We find that in a
semi-infinite geometry, when the surface also undergoes a cyclic
evolution, the pumping is obstructed by the appearence of ``surface
Weyl points'' that transfer quanta of Berry-curvature flux between the
valence and conduction bands.  Some additional features of the model
are discussed in two Appendices.  Appendix~\ref{app:mirror} deals with
the quantization of the CSA coupling by mirror symmetry at isolated
points along the pumping cycle, and in Appendix~\ref{app:2-param} we
map the second Chern number over an augmented parameter space.

\section{CSA coupling in the hybrid Wannier representation}

\subsection{Definitions and conventions}
\label{sec:defs}

The CSA coupling strength $\theta_{\rm CS}$ of an infinite 3D crystal
was defined in \eq{theta_3d}. In the remainder of the paper we will
change notation and denote it as $\tthree$ instead. The reason is that
in order to establish the precise relation between the bulk CSA
coupling and the observable magnetoelectric response of a bounded
macroscopic sample, we also want to consider systems of reduced
dimensionality $d$, namely slabs ($d\!=\!2$) and crystallites
($d\!=\!0$), and define for them corresponding quantities $\ttwo$ and
$\tzero$.  In each case $\theta_d$ is defined as the extensive CSA
coupling with units of volume, divided by the cell ``volume'' along
the periodic directions, so that $\theta_d$ has units of $[L]^{3-d}$:
$\tzero$ grows with the volume of the crystallite, $\ttwo$ with the
slab thickness, and $\tthree\equiv\theta_{\rm CS}$ is an intensive
bulk quantity.

We assume that the crystal lattice has monoclinic or higher symmetry.
The lattice vectors form a prism whose base of area $A_{\rm cell}$
spanned by $\ba_1$ and $\ba_2$ lies in the $(x,y)$ plane, and
$\ba_3=c\hat{\bm z}$ is the unique axis.  The Bloch functions are
normalized to one unit cell, $\int_{V_{\rm cell}}d\br\vert\psi_{\bk
  n}(\br)\vert^2=1$ with $V_{\rm cell}=cA_{\rm cell}$. In the HWF
representation, the $\hat{\bm z}$ direction is treated differently
from the $(x,y)$ plane. Defining $\kpar =(k_x,k_y)$ and writing as
$\langle f\vert g\rangle$ the integral over all $z$ and over one cell
on the basal plane, the orthonormality relation for the Bloch states
reads
\beq
\label{eq:orthonormality}
\ip{\psi_{(\kpar,k_z)n}}{\psi_{(\kpar',k_z')m}}=
N_z\delta_{\kpar\kpar'}\delta_{k_zk_z'}\delta_{nm},
\eeq
where $N_z$ is the number of cells along $z$, and
$N_z\delta_{k_zk_z'}$ becomes $(2\pi/c)\delta(k_z-k_z')$ when
$N_z\rightarrow\infty$.  Choosing as the BZ a prism of height $2\pi/c$
and base lying on the $(k_x,k_y)$ plane, the bulk HWFs
$\ket{\hwf_{\kpar ln}}$ and their cell-periodic parts $\ket{h_{\kpar
    ln}}$ are defined as
\begin{subequations}
\label{eq:hwf}
\begin{align}
\label{eq:hwf-bloch}
\ket{\hwf_{\kpar ln}}&=\frac{c}{2\pi}\int_{0}^{2\pi/c} dk_z\,
e^{-ik_zlc}\,\ket{\psi_{(\kpar,k_z)n}}\\
\label{eq:hwf-per}
\ket{h_{\kpar ln}}&=e^{-i\kpar\cdot\br}\,\ket{\hwf_{\kpar ln}},
\end{align}
\end{subequations}
where $l$ labels the cells in the $z$ direction. The HWFs are
localized (Wannier-like) along $z$, but remain extended (Bloch-like)
in the other two directions.  Using \eq{orthonormality} we obtain
$\ip{\hwf_{\kpar ln}}{\hwf_{\kpar' l'm}}=\delta_{\kpar\kpar'}\delta_{ll'}\delta_{nm}$.

We are interested in constructing HWFs that span the group of $M$
valence bands. The $U(M)$ gauge freedom of \eq{gauge-transformation}
in defining the Bloch states can be employed to make the HWFs
maximally-localized along $z$, in the sense of
Ref.~\onlinecite{marzari-prb97}. The charge centers of these maximally
localized HWFs
\beq
\label{eq:z-sheet}
z_{\kpar ln}=\me{\hwf_{\kpar ln}}{z}{\hwf_{\kpar ln}}=z_{\kpar 0n}+lc 
\eeq
form 2D sheets over the projected BZ.\cite{taherinejad-prb14} Dropping
subscripts $\kpar$ for brevity, Berry connection and Berry curvature
matrices can be defined over these sheets as
\begin{subequations}
\label{eq:berry-bulk}
\begin{align}
\label{eq:connection-bulk}
A^i_{ln,l'm}&=i\ip{h_{ln}}{\partial_{k_i} h_{l'm}}=A^i_{0n,(l'-l)m}\\
\label{eq:curv-bulk}
\Omega^{ij}_{ln,l'm}&=\partial_{k_i} A^j_{ln,l'm}-\partial_{k_j} A^i_{ln,l'm}
=\Omega^{ij}_{0n,(l'-l)m}
\end{align}
\end{subequations}
where $i,j=x,y$.

Assuming that the Wannier sheets do not touch (i.e., there are no
degeneracies in the $z_{ln}$ anywhere in the 2D BZ), each sheet is a
closed 2D manifold, so that the integral of its Berry curvature (the
Berry flux through the 2D BZ) is quantized to $2\pi$ times an integer
by the Chern theorem.  These Chern numbers
\beq
\label{eq:chern}
C_{ln}=\frac{1}{2\pi}\int d\kpar\,\Omega^{xy}_{ln,ln}
\eeq
are clearly independent of the layer index, $C_{ln}=C_{0n}$.

We shall encounter situations in Sec.~\ref{sec:haldane} where pairs of
Wannier sheets touch at isolated points in the 2D BZ as a parameter
$\cycle$ in the Hamiltonian is varied. When that happens, the Chern
numbers of the two sheets change by equal and opposite amounts
$\pm\chi$, where $\chi$ is the chiral charge of the degeneracy point
(``Weyl point'') in the 3D parameter space $(k_x,k_y,\cycle)$.

\subsection{CSA coupling $\tthree$ of a bulk crystal}
\label{sec:hwf}

We begin by summarizing the results of
Ref.~\onlinecite{taherinejad-prl15}, where the bulk CSA coupling was
expressed in terms of maximally-localized HWFs as\footnote{The
  expression for $\theta_{\Delta xy}$ in Eq.~(12) of
  Ref.~\onlinecite{taherinejad-prl15} has a typo in it: a minus sign
  is missing.}
\begin{subequations}
\label{eq:theta_3d_tot}
\begin{align}
  \tthree&=\theta_{z\Omega}+\theta_{\Delta xy}\label{eq:theta_3d_final}\\
  \theta_{z\Omega}&=-\frac{1}{c}\int d\kpar\sum_{n}
  z_{0n}\Omega^{xy}_{0n,0n}\label{eq:zomega}\\
  \theta_{\Delta xy}&=-\frac{i}{c}\int d\kpar\sum_{lnm}
  \left(z_{lm}-z_{0n}\right)A^x_{0n,lm}A^y_{lm,0n}.\label{eq:dxy}
\end{align}
\end{subequations}
(Henceforth summations over indices $l$, $m$, and $n$ run over all
occupied orbitals, unless stated otherwise.)

The maximally-localized HWF gauge is unique, except for (i) a $U(1)$
gauge freedom with respect to $\kpar$ on each sheet, and (ii) the
residual freedom to choose which Wannier sheets belong to the home
cell $l=0$.  Regarding (i), it is straightforward to verify that both
terms in \eq{theta_3d_tot} are gauge-invariant in this sense (for
$\theta_{\Delta xy}$ this follows because only the diagonal elements
of $A^i_{0n,lm}$ are affected).  Regarding (ii), if a different choice
is made such that entire sheets $z_{ln}$ get shifted by $c$ for
some~$n$, $\theta_{z\Omega}$ changes by $-2\pi C_{0n}$, while
$\theta_{\Delta xy}$ is unaffected.  Thus, the $\theta_{z\Omega}$ term
is the only one that has a potential $2\pi$ ambiguity.

In Ref.~\onlinecite{taherinejad-prl15}, \eq{theta_3d_tot} was derived
starting from \eq{theta_3d}. One problem with this approach is that
\eq{theta_3d} is written in terms of a smooth and periodic gauge, a
requirement that, applied to the maximally-localized HWF gauge, is
incompatible with the existence of Wannier sheets having nonzero Chern
numbers.  Thus, strictly speaking \eq{theta_3d_tot} has only been
proven to be valid for crystals in which all the $C_{0n}$ vanish.  On
the other hand, it was shown in the same work that nonzero $C_{0n}$
values must occur along any cycle that pumps a quantum of CSA
coupling.

In the following, we shall take a different route to arrive at
\eq{theta_3d_tot}, proving the surface theorem along the way. Instead
of working from the outset with a bulk crystal and using
\eq{theta_3d}, we start from the CSA coupling of a finite crystallite,
which is given by \eq{theta_0d} below without any $2\pi$ ambiguity.
By carefully taking the thermodynamic limit, first in two directions
(slab geometry) and finally in the third, we will demonstrate that
\eq{theta_3d_tot} remains valid even when some of the Wannier sheets
have nonzero Chern numbers, provided that the unit-cell sum of those
numbers vanishes:
\beq
\label{eq:chern-sum}
\sum_n C_{0n}=0.
\eeq
This condition, needed to ensure that the nonquantized part of
$\tthree$ is independent of the choice of origin for the $z$~axis, is
equivalent to the statement that the Chern index along $z$ of the
valence-band manifold vanishes.\cite{taherinejad-prl15} In fact, in
the present work we limit ourselves to crystals in which all three
Chern indices vanish, in order to avoid potential subtleties
associated with bulk quantum anomalous Hall behavior.

\subsection{CSA coupling $\tzero$ of a finite crystallite}
\label{sec:crystallite}

The extensive CSA coupling of a bounded electron system such as
a crystallite is given by
\begin{equation}\label{eq:theta_0d}
 \tzero=-8\pi^2\text{ImTr}[PxPyPz],
\end{equation}
where $P$, the projection operator onto the occupied subspace in the
ground state, is expressed in the energy eigenstate representation as
$P=\sum_n\ket{\psi_n}\bra{\psi_n}$.

\equ{theta_0d} was obtained in Ref.~\onlinecite{malashevich-njp10}
starting from the basic definition
\beq
\label{eq:m-E}
{\bm m}({\bm E})=-\frac{e}{2}{\rm Tr}
\left[P({\bm E}){\bm r}\times{\bm v}\right]
\eeq
of the orbital magnetic moment of a bounded sample in a finite
electric field, with $P({\bm E})$ the projection operator onto the
{\it field-polarized} occupied states. After some manipulations,
\eq{m-E} was decomposed as~\cite{malashevich-njp10}
\beq
{\bm m}({\bm E})=\wt{\bm m}_{\rm LC}({\bm E})+\wt{\bm
  m}_{\rm IC}({\bm E})+{\bm
  m}_{\rm CS}({\bm E}).
\eeq
At ${\bm E}=0$, only the first two terms survive (they are known as
the ``local circulation'' and the ``itinerant
circulation''\cite{ceresoli-prb06}). At ${\bm E}\not= 0$ those terms
depend on ${\bm E}$ only through $P({\bm E})$, and their contribution
to the linear change in ${\bm m}$ induced by ${\bm E}$ is in general
not parallel to ${\bm E}$.  In contrast, the third term is purely
isotropic and has an {\it explicit} linear dependence on ${\bm E}$. It
takes the form
\beq
\label{eq:m-CS-E}
{\bm m}_{\rm CS}({\bm E})
=\frac{e^2}{h}\frac{\theta_0({\bm E})}{2\pi}{\bm E},
\eeq
with $\theta_0({\bm E})$ given in terms of $P({\bm E})$ by
\eq{theta_0d}.  At linear order in the field, the quantity
$\theta_0({\bm E})$ in \eq{m-CS-E} can be replaced with $\theta_0({\bm
  E}=0)$, i.e., the {\it ground-state} quantity~$\theta_0$ as
originally defined by \eq{theta_0d}.

According to the analytical derivations and numerical tests carried
out in Ref.~\onlinecite{malashevich-njp10}, in the thermodynamic limit
the quantity $\theta_0/V$ reduces to the bulk CSA coupling $\theta_3$
defined by \eq{theta_3d}. However, the fact that $\theta_3$ suffers
from a $2\pi$ indeterminacy while $\theta_0/V$ does not was not
examined further in that work.  Because \eq{theta_0d} does not carry a
quantum of uncertainty, and is free from the subtleties associated
with the use of periodic boundary conditions, we take it as the more
fundamental definition of the CSA coupling, from which we will derive
expressions for $\ttwo$ and $\tthree$ in the HWF representation. The
correctness of those expressions will be checked numerically against
results obtained directly from \eq{theta_0d}.

We begin by constructing for the crystallite a set of occupied
orbitals that are maximally-localized along $z$. This can be done by
diagonalizing the operator $PzP$,\cite{marzari-prb97}
\beq
\label{eq:PzP-0d}
PzP|\varphi_n\rangle=z_n|\varphi_n\rangle.
\eeq
In this representation, \eq{theta_0d} reads
\beq
\label{eq:theta_0d_chern_P}
\tzero =-8\pi^2\sum_{n}z_n\text{Im}\me{\varphi_n}{xPy}{\varphi_n}
\eeq
or equivalently
\beq
\label{eq:theta_0d_chern}
\tzero
=8\pi^2\sum_{n}z_n\text{Im}\me{\varphi_n}{xQy}{\varphi_n},
\eeq
where $Q=1-P$ and we used $\text{Im}\me{\varphi_n}{ xy}{\varphi_n}=0$.
We shall prefer \eq{theta_0d_chern} for reasons that will become clear
in the next subsection.

Before switching to a slab geometry, we note that any well-defined
intrinsic property must remain invariant under a rigid translation of
the sample, such as
\beq
\label{eq:z-shift}
z\rightarrow z + \Delta z.
\eeq
Under this transformation, \eq{theta_0d_chern} changes by
\begin{align}\label{eq:delta_theta_0d_chern}
\Delta\tzero
=8\pi^2\Delta z\text{ImTr}\{PxQy\}.
\end{align}
It is straightforward to verify that $\Delta\tzero=0$ since
$\text{ImTr}\{PxPy\}=\text{Tr}\{[Px,Py]\}/2i$, and the trace of a
commutator over a finite-dimensional Hilbert space vanishes.

\subsection{CSA coupling $\ttwo$ of a slab}
\label{sec:slab}

Consider an insulating slab of thickness $L_z=N_z c$.  We will
eventually take the limit $N_z\rightarrow\infty$, but at this stage
$N_z$ can be any positive integer.  We wish to find an expression for
~$\ttwo$, the CSA coupling per unit area, starting from
\eq{theta_0d_chern}.  Imagine cutting from the slab a crystallite
containing $N\times N$ two-dimensional primitive cells with a net CSA
coupling $\tzero$.  By the definition of~$\ttwo$, we expect that
\begin{align}\label{eq:0d-2d}
\frac{\tzero}{N^2 A_{\rm cell}}
\rightarrow\ttwo\,\,\,\,\,\,\text{for } N\rightarrow\infty.
\end{align}
As we will see, this expectation is fulfilled by our expression for
$\ttwo$, with the proviso that the net Chern number of the slab
vanishes; otherwise $\ttwo$ is ill-defined, with its value depending
on the choice of origin for the $z$~axis. Such pathological behavior
has to do with extensive contributions to $\tzero$ from the sample
edges.  We shall return to this subtle point at the end of the
section, starting with \eq{delta_theta_2d}.

To proceed we assume a slab with unreconstructed surfaces, so that the
energy eigenstates are Bloch-like along the in-plane directions;
denoting them by $|\psi_{\kpar n}\rangle$ and introducing the
ground-state projector
\begin{align}\label{eq:P-slab}
  P=\sum_{\kpar}P_{\kpar}, \qquad P_{\kpar}=\sum_n\,
  \ket{\psi_{\kpar n}}\bra{\psi_{\kpar n}},
\end{align}
we can construct a new set of occupied orbitals by diagonalizing
$P_{\kpar}zP_{\kpar}$ for each $\kpar$:
\beq
\label{eq:PzP-2d}
P_{\kpar}zP_{\kpar}\,\ket{\hwf_{\kpar n}}=z_{\kpar n}\,\ket{\hwf_{\kpar n}}.
\eeq
These are the maximally-localized HWFs of the slab. For now we assume
isolated Wannier sheets $z_{\kpar n}$ with well-defined Chern numbers
$C_n=(1/2\pi)\int d\kpar\Omega^{xy}_{\kpar nn}$; the role of
degeneracies will be considered later.

Comparison with \eq{PzP-0d} shows that for $N\rightarrow\infty$ the
orbitals $\ket{\hwf_{\kpar n}}$ and $\ket{\varphi_n}$ only differ in
the choice of in-plane boundary conditions (periodic versus
open). Thus, if edge contributions to $\tzero$ are unimportant we can
replace $\ket{\varphi_n}$ with $\ket{\hwf_{\kpar n}}$ in
\eq{theta_0d_chern}, and using the definition~(\ref{eq:0d-2d}) of
$\ttwo$ we arrive at
\beq
\label{eq:theta_2d}
\ttwo=\int d\kpar\sum_n\,
z_{\kpar n}\,2\text{Im}\me{\hwf_{\kpar n}}{xQy}{\hwf_{\kpar n}}.
\eeq
The presence of the nonperiodic coordinate operators $x$ and $y$ is
not problematic, because they appear in the lattice-periodic
combinations $PxQ$ and $QyP$.

\equ{theta_2d} is valid in the maximally-localized HWF gauge where
$Z_{\kpar nm}=\me{\hwf_{\kpar n}}{z}{\hwf_{\kpar m}}$ is a diagonal
matrix: $Z_{\kpar nm}=z_{\kpar n}\delta_{nm}$. In a generic gauge,
\eq{theta_2d} must be written in matrix form as
\begin{subequations}
\begin{align}
\label{eq:theta_2d_trace_a}
\ttwo&=-\int d\kpar\sum_{mn}\,Z_{\kpar mn}\widetilde{\Omega}^{xy}_{\kpar nm},\\
\label{eq:cov-curv-a}
\widetilde{\Omega}^{xy}_{\kpar nm}
&=-2\text{Im}\me{\hwf_{\kpar n}}{xQy}{\hwf_{\kpar m}}.
\end{align}
\end{subequations}
This expression is clearly gauge-invariant, because it is the trace of
the product of two gauge-covariant matrices (matrices that change as
$M_{\kpar}\rightarrow U^\dagger_{\kpar} M_{\kpar} U_{\kpar}$ under a
gauge transformation $\ket{\hwf_{\kpar n}}\rightarrow
\sum_m\ket{\hwf_{\kpar m}}U_{\kpar mn}$).

To proceed we switch momentarily to the Hamiltonian gauge, where the
orbitals $\ket{\hwf_{\kpar n}}$ coincide with the energy eigenstates
$\ket{\psi_{\kpar n}}$. Writing $Q=\sum_{\kpar'}\sum_l^{\rm
  empty}\ket{\psi_{\kpar' l}}\bra{\psi_{\kpar' l}}$ in
\eq{cov-curv-a}, we see that it only involves off-diagonal matrix
elements of the coordinate operators along the periodic directions.
Defining the cell-periodic part $\ket{u_{\kpar
    n}}=e^{-i\kpar\cdot\br}\ket{\psi_{\kpar n}}$ of an eigenstate and
using the relation\cite{blount-ssp62}
\beq
\me{\psi_{\kpar n}}{x}{\psi_{\kpar' m}}=
i\ip{u_{\kpar n}}{\partial_{k_x} u_{\kpar m}}\delta_{\kpar\kpar'}
\eeq
valid for $n\not=m$ in the Hamiltonian gauge, we find
\bea
\label{eq:cov-curv}
\widetilde{\Omega}^{xy}_{\kpar nm}&=&-2\text{Im}
\me{\partial_{k_x}u_{\kpar n}}{Q_{\kpar}}{\partial_{k_y}u_{\kpar m}}\nn
&=&\Omega^{xy}_{\kpar nm}
-i[A^x_{\kpar},A^y_{\kpar}]_{nm},
\eea
where $Q_{\kpar}=\sum_l^{\rm empty}\ket{u_{\kpar l}}\bra{u_{\kpar
    l}}$.  In the second line the completeness relation was used to
obtain an expression containing the Berry connection and Berry
curvature matrices $A^x_{\kpar nm}$ and $\Omega^{xy}_{\kpar nm}$ for
the occupied states, defined in terms of the slab orbitals by
equations like \eq{berry-bulk} but with the replacements
$ln\rightarrow n$ and $l'm\rightarrow m$.  The matrix in \eq{cov-curv}
is known as the non-Abelian Berry curvature,\cite{mead-rmp92} and it
follows from its gauge covariance that \eq{theta_2d_trace_a} can be
written in any HWF gauge as
\beq
\label{eq:theta_2d_gauge_inv}
\ttwo=-\int d\kpar\,
\text{Tr}\left(Z_{\kpar}\widetilde{\Omega}^{xy}_{\kpar}\right),
\eeq
where \eq{cov-curv-a} for $\widetilde{\Omega}^{xy}_{\kpar}$ has been
replaced with \eq{cov-curv}.  Returning to the maximally-localized HWF
gauge,
\beq
\label{eq:theta_2d_per}
\ttwo=-\int d\kpar\sum_n\,z_{\kpar n}\widetilde{\Omega}^{xy}_{\kpar nn}. 
\eeq
In Sec.~\ref{sec:proof} we will recast this expression in a form
similar to \eq{theta_3d_tot} for~$\tthree$, but for the moment (and
also for numerics, see Sec.~\ref{sec:results-bilayer})
\eq{theta_2d_per} is more convenient.

Equations~(\ref{eq:theta_2d_gauge_inv}) and~(\ref{eq:theta_2d_per})
constitute the central result of this section. We emphasize that
nowhere in their derivations was it assumed that the Chern numbers
$C_n$ of the individual Wannier sheets must vanish.

Let us conclude with a discussion of the conditions needed in order
for $\ttwo$ to be a well-defined quantity.  Under the transformation
(\ref{eq:z-shift}), \eq{theta_2d_per} changes by
\beq
\label{eq:delta_theta_2d}
\Delta\ttwo=-2\pi C_{\rm slab}\Delta z,
\eeq
where $C_{\rm slab}=\sum_nC_n$ is the Chern number of the entire slab,
and we used the fact that the trace of the commutator in \eq{cov-curv}
vanishes.  Contrary to $\Delta\tzero$ which is guaranteed to vanish
for any crystallite, $\Delta\theta_2$ is nonzero for quantum anomalous
Hall slabs where $C_{\rm slab}$ is a nonzero integer.  Clearly,
$\ttwo$ is ill-defined in such systems.

How can one reconcile $\Delta\ttwo\not= 0$ for a quantum anomalous
Hall slab with $\Delta\tzero=0$ for a crystallite cut from the same
slab?  To understand this result, we invoke the definition
\beq
C(\br)=-4\pi\text{Im}\me{\br}{PxQy}{\br}
\eeq
of the ``local Chern marker''\cite{bianco-prb11} to rewrite
\eq{delta_theta_0d_chern} as
\beq
\Delta\tzero=-2\pi\left[\int d\br\,C(\br)\right]\Delta z,
\eeq
which can be directly compared with \eq{delta_theta_2d} for
$\Delta\ttwo$.  Whenever $C_{\rm slab}\not= 0$, $\int d\br\,C(\br)$
vanishes in a nontrivial manner, with extensive contributions from the
edges of the crystallite exactly cancelling those from the
interior.\cite{bianco-prb11} When we switched from open to periodic
boundary conditions along $x$ and $y$ to obtain \eq{theta_2d}, the
edges were supressed.  Clearly, that step relied on the absense of
extensive edge contributions to $\int d\br\,C(\br)$, or equivalently,
\beq
\label{eq:C-slab}
C_{\rm slab}=0.
\eeq

\subsection{Surface theorem for the CSA coupling}
\label{sec:surf-thm}

\subsubsection{Statement of the theorem}

Armed with \eq{theta_2d_per} for the CSA coupling of an insulating
slab, let us relate it to the bulk coupling~$\tthree$ by considering
the thermodynamic limit. Defining the dimensionless slab CSA coupling
\beq
\label{eq:theta-slab}
\theta_{\rm slab}\equiv\frac{\ttwo}{N_z c}
\eeq
and reasoning by analogy with \eq{0d-2d}, one might expect that
\beq
\theta_{\rm slab}
\stackrel{?}{\rightarrow}\tthree
\,\,\,\,\,\,\text{for } N_z\rightarrow\infty.
\eeq
This cannot be quite correct, because $\tthree$ is only defined modulo
$2\pi$ whereas $\theta_{\rm slab}$ is uniquely defined. Turning this
observation around, it should be possible to resolve the quantum of
ambiguity in $\tthree$ by isolating quantized surfacelike
contributions to $\theta_{\rm slab}$.

\begin{figure}[tb]
\center
\includegraphics[width=1.0\columnwidth]{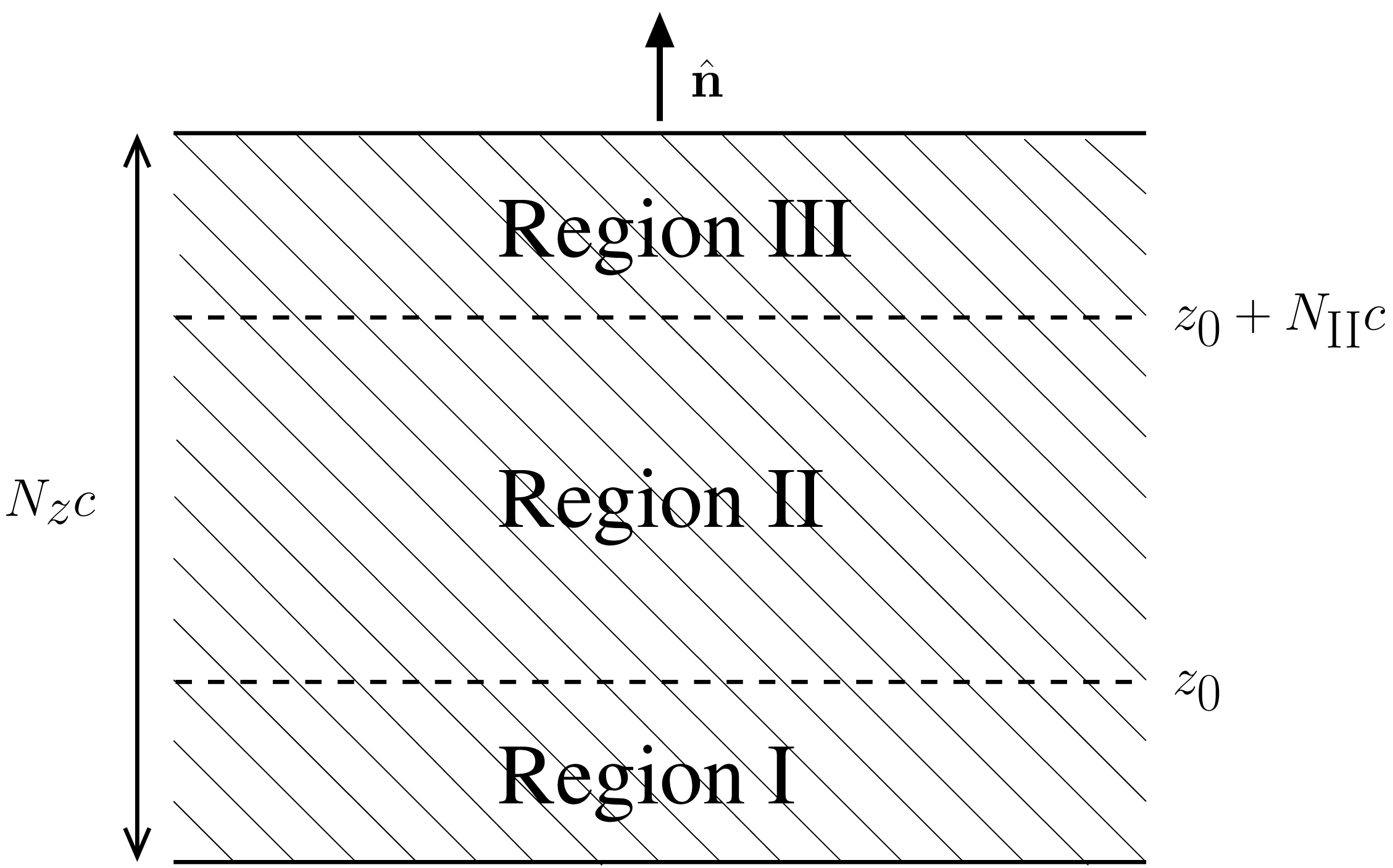}
\caption{Sketch of the slab configuration discussed in the text
  (adapted from Ref.~\onlinecite{vanderbilt-prb93}). The HWFs are
  localized in the surface-normal direction
  $\hat{\bm n}=\hat{\bm z}$.}
\label{fig:slab-sketch}
\end{figure}

It is useful to divide the slab conceptually into three regions
(Fig.~\ref{fig:slab-sketch}): a bottom surface region~I, an interior
region~II containing precisely $N_{\rm II}$ bulk unit cells, and a top
surface region~III.  The boundaries are chosen to respect the HWF
sheets, so that each sheet belongs uniquely to one region.  We assume
that the surface regions occupy nonextensive fractions of the volume
of the slab, but are nonetheless sufficiently thick to ensure that the
cuts $z_0$ and $z_0+N_{\rm II}c$ fall in bulklike regions.

We will prove the following assertion, valid when the Fermi level lies
in a gap common to both the bulk and surface bands:
\beq
\label{eq:surf-thm}
\theta_{\rm slab}
\rightarrow\tthree-2\pi C_{\rm III} \,\,\,\,\,\,\text{for }
N_z\rightarrow\infty, 
\eeq
with $\tthree$ expressed in terms of the bulklike HWFs in region~II by
\eq{theta_3d_tot}, and $C_{\rm III}=\sum_{n\in {\rm III}}C_n$ the
total integer Chern number of the HWFs ascribed to the top surface
region.  The Chern indices of the valence-band manifold are assumed to
vanish in all three lattice directions, which implies [see
\eq{chern-sum}]
\beq
C_{\rm II}=\sum_{n\in {\rm II}}C_n=N_{\rm II}\sum_nC_{0n}=0.
\eeq
From \eq{C-slab} we also require $C_{\rm I}+C_{\rm II}+C_{\rm III}=0$,
so that
\beq
\label{eq:C_A-C_B}
C_{\rm I}+C_{\rm III}=0.
\eeq

Once the assignment of the slab HWF sheets to the three regions has
been decided, the two terms on the right-hand-side of \eq{surf-thm}
become uniquely defined. That is, \eq{theta_3d_tot} must be evaluated
using the choice of unit cell consistent with the $N_{\rm II}$ bulk
cells in region~II, and the leftover sheets then contribute to
$C_{\rm I}$ and $C_{\rm III}$. If a different assignment is made, an
integer multiple of $2\pi$ may get transferred between the two
contributions to \eq{surf-thm}.  We will assume that a specific
assignment has been made that satisfies two rules:
\begin{enumerate}

\item Sheets belonging to a given region have sequential indices $n$,
  with $z_{\kpar n}\leq z_{\kpar,n+1}$.

\item The last sheet belonging to region~I does not touch the first
  sheet belonging to region~II.

\end{enumerate}
Because region~II comprises an integer number of bulk cells, the
second rule also implies that the topmost sheet in region~II does not
touch the first sheet in region~III, guaranteeing that the Chern
numbers of the two surface regions are well defined.  This should be
possible to arrange in most cases --~even at critical parameter values
where two Wannier sheets touch at a high symmetry point~-- by a
judicious choice of the first sheet in region~II. (One exception is a
$\Z2$-odd topological insulator protected by $\trev$ symmetry, where
there are no gaps between bulk Wannier sheets: see, for example, the
third panel in Fig.~1 of Ref.~\onlinecite{taherinejad-prl15}.  Before
applying the current analysis to such a system, one would have to
break $\trev$ symmetry slightly to gap the spectrum.)

\equ{surf-thm} can be understood as follows.  The assignment of the
HWFs to the three regions decides which ones belong to a given unit
cell in the interior region.  This removes the only gauge freedom that
is capable of affecting the branch choice for the phase angle
$\tthree$ in \eq{theta_3d_tot}.  The extra term $-2\pi C_{\rm III}$
gathers the contributions to the slab CSA coupling from any leftover
sheets after tiling the bulk cell towards the top surface.  This is
analogous to the electric polarization of a
slab,\cite{vanderbilt-prb93} where the relevant quantized quantity
carried by the Wannier sheets is the charge rather than the Chern
number. (One difference is that Wannier sheets can have different
Chern numbers, while every sheet carries the same charge $-e$.)

\subsubsection{Proof of the theorem }
\label{sec:proof}

In order to establish \eq{surf-thm}, we start from the
left-hand-side. Using \eqs{theta_2d_per}{cov-curv} and exchanging
indices $n$ and $m$ in one term we find, approximating the slab width
$N_z c$ by the width $N_{\rm II}c$ of the interior region,
\begin{subequations}
\begin{align}
\theta_{\rm slab}&=-\frac{1}{c}\int d\kpar\,\vartheta_{\rm slab}\\
\label{eq:vartheta-slab}
\vartheta_{\rm slab}&=
\frac{1}{N_{\rm II}}
\sum_n\left[
  z_n\Omega^{xy}_{nn}+i\sum_m(z_m-z_n)A^x_{nm}A^y_{mn}
\right],
\end{align}
\end{subequations}
where the label $\kpar$ has been dropped from the integrand.  (Here
and for the rest of this section, it is understood that certain
equalities only hold exactly for $N_{\rm II}\rightarrow\infty$.) The
similarity to \eq{theta_3d_tot} for $\tthree$ is apparent.

Next we decompose $\theta_{\rm slab}$ and $\vartheta_{\rm slab}$ into
contributions from each region,
\beq
\label{eq:theta-A+I+B}
\theta_{\rm slab}=
\theta_{\rm I}+\theta_{\rm II}+\theta_{\rm III},
\eeq
by restricting the summation over $n$ in \eq{vartheta-slab} to the
sheets assigned to that region (the index $m$ is still allowed to run
over all Wannier sheets in the slab).

Let us start with the top surface region.  The contribution to
$\vartheta_{\rm III}$ from the second term in \eq{vartheta-slab}
vanishes for a thick slab, because (i) that term involves the {\it
  relative} coordinate $z_m-z_n$, (ii) $A^x_{nm}A^y_{mn}$ drops
exponentially to zero when $|z_m-z_n|^2$ is much larger than the HWF
spread $\langle z_n\rangle^2-\langle z_n\rangle^2$, and (iii)
region~$A$ is nonextensive. We are left with the first term in
\eq{vartheta-slab}, which involves the {\it absolute} coordinate
$z_n$. In the limit $N_{\rm II}\rightarrow \infty$ we can set
$z_n\rightarrow z_0+N_{\rm II}c$ for $n\in {\rm III}$ (see
Fig.~\ref{fig:slab-sketch}) to obtain
\beq
\theta_{\rm III}
=-2\pi\left( 1+\frac{z_0}{N_{\rm II}c}\right)C_{\rm III}.
\eeq
Similarly, $\theta_{\rm I}=-2\pi(z_0/N_{\rm II}c)C_{\rm I}$. Adding
the two together and using \eq{C_A-C_B} yields the net surface
contribution
\beq
\label{eq:theta-A+B}
\theta_{\rm surf}
=\theta_{\rm I}+\theta_{\rm III}=-2\pi C_{\rm III},
\eeq
which is origin independent and quantized in units of $2\pi$.

Now we turn to the contribution from the interior region.  The
quantity $\vartheta_{\rm II}$ can be approximated as
\beq
\label{eq:vartheta-I}
\vartheta_{\rm II}=\frac{1}{N_{\rm II}}
\left[
  \sum_{n\in {\rm II}}z_n\Omega^{xy}_{nn}
  +i\sum_{nm\in {\rm II}}(z_m-z_n)A^x_{nm}A^y_{mn}
\right].
\eeq
By also restricting the summation over $m$ to region~II, we have
dropped nonextensive contributions.  Since that region is bulklike, we
can switch to the notation of Sec.~\ref{sec:hwf}.  Replacing
$n\rightarrow ln$ and $m\rightarrow l'm$ and invoking
\eqs{z-sheet}{berry-bulk}, the first term in \eq{vartheta-I} becomes
\beq
\vartheta_{z\Omega}=\sum_n\,z_{0n}\Omega^{xy}_{0n,0n}
\eeq
[we have dropped a term
$(1/N_{\rm II})\sum_l(lc)\sum_n\Omega^{xy}_{0n,0n}$ that vanishes upon
integration over $\kpar$ by virtue of \eq{chern-sum}].  As for the
second term, it becomes
\beq
\vartheta_{\Delta xy}=
i\sum_{lnm}\,(z_{lm}-z_{0n})A^x_{0n,lm}A^y_{lm,0n}.
\eeq
Adding the two and integrating over $\kpar$ we find, comparing with
\eq{theta_3d_tot},
\beq
\label{eq:theta-I}
\theta_{\rm II}=\theta_{z\Omega}+\theta_{\Delta xy}=\tthree.
\eeq
Combining Eqs.~(\ref{eq:theta-A+I+B}), (\ref{eq:theta-A+B}),
and~(\ref{eq:theta-I}) we arrive at \eq{surf-thm}, completing the
proof of the surface theorem. Note that condition~(\ref{eq:chern-sum})
insures that the two terms on the right-hand-side of \eq{surf-thm} are
separately origin independent.

\section{Layered Haldane model}
\label{sec:haldane}

In this section we illustrate the preceeding discussion with a
concrete example: a 3D tight-binding model of an insulator that pumps
a quantum of CSA coupling during a cyclic adiabatic evolution.

The model consists of a stack of half-filled
Haldane-model~\cite{haldane-prl88} (``haldanium'') layers placed
exactly on top of each other, that are allowed to interact via
interlayer hoppings.  The on-site energies are equipped with a
dependence on a cyclic tuning parameter $\cycle$, such that if the
layers were decoupled, their Chern numbers would either be all zero or
alternate between $+1$ and $-1$ from one layer to the next, depending
on the value of~$\cycle$. (An isolated layer with nonzero Chern number
$C$ in the valence band is a quantum anomalous Hall insulator with AHC
$\sigma_{yx}=Ce^2/h$.)

For decoupled layers, the system would pass through metallic points at
critical $\cycle$ values where the Chern numbers change. By
introducing $\cycle$-dependent interlayer couplings, it is possible to
keep the bulk insulating throughout the cycle.  The band touchings at
the critical $\cycle$ values are then replaced with touchings between
Wannier sheets residing mostly on individual layers, accompanied by a
transfer of Chern number (the total Chern number of the valence bands
vanishes in all three directions).  Each Wannier sheet participates in
two touching events per cycle, one with the sheet below and another
with the sheet above, resulting in the net transfer across a unit cell
of a Chern number of $-1$.

The numerical results presented below were obtained using the {\tt
  PythTB} code package.\footnote{Available at
  \url{http://physics.rutgers.edu/pythtb/}.}

\subsection{Cyclic evolution of a bilayer model}
\label{sec:results-bilayer}

We begin by considering a model consisting of only two coupled
haldanium layers. Although there is no CSA pumping in this model, it
serves to illustrates the elementary sheet-touching events during a
cyclic evolution, and provides the building blocks for the full 3D
model.

\begin{figure}
\begin{center}
{\bf (a)} \includegraphics[width=4cm]{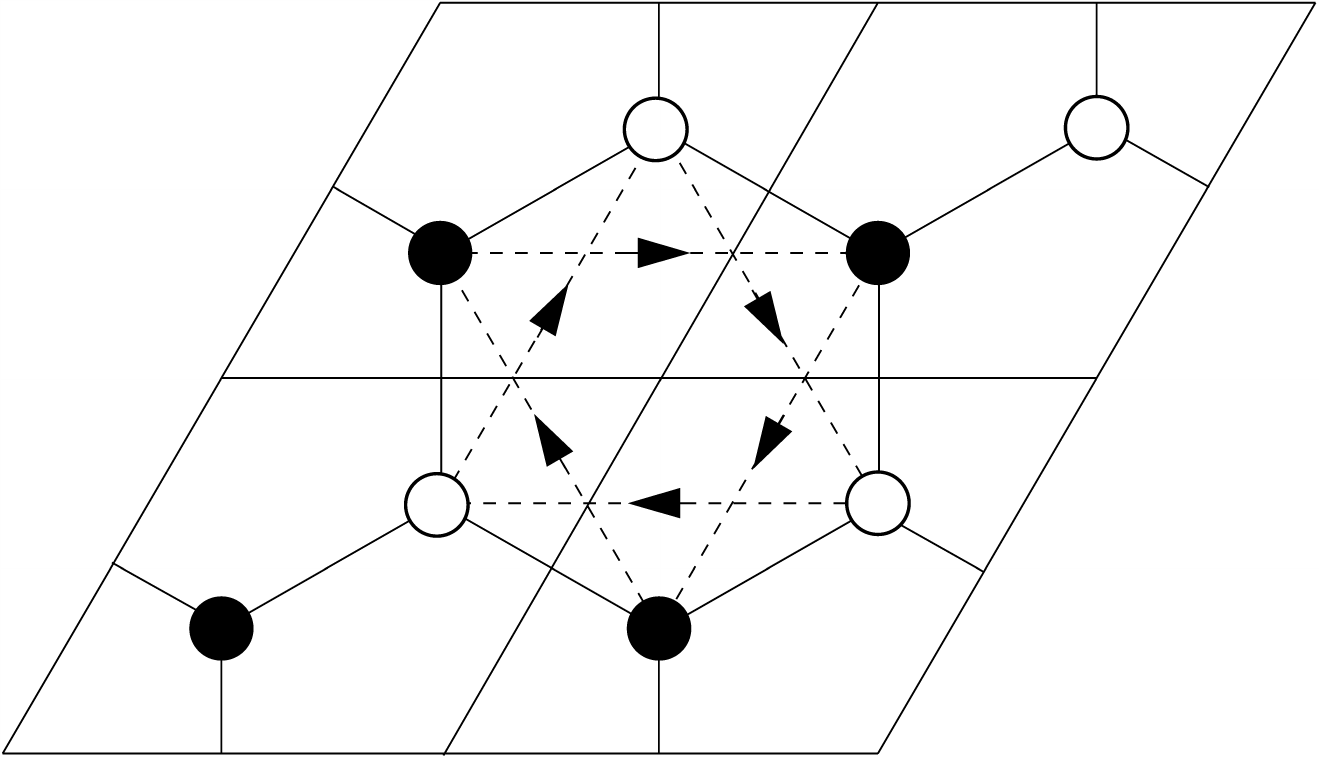}\quad
{\bf (b)} \includegraphics[width=2.7cm]{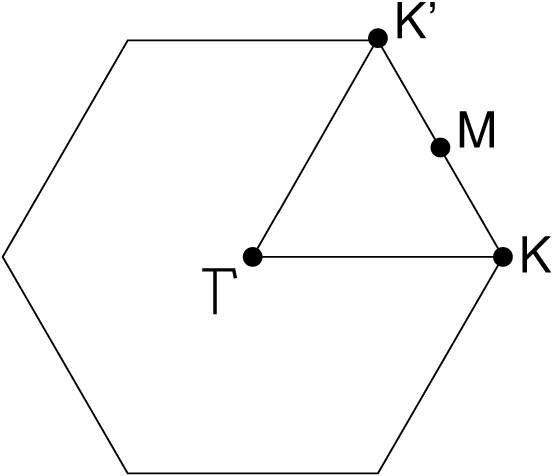}
\end{center}
\caption{\label{fig:model} (a) Four unit cells of an isolated
  haldanium layer. Sites on the $A$ ($B $) sublattices are marked with
  open (filled) circles, and the arrows indicate the directions of the
  second nearest neighbor hoppings in \eq{H-monolayer} with amplitudes
  $i(-1)^lt_2$. (b) First Brillouin zone, with high-symmetry points
  marked.  (Reproduced from Ref.~\onlinecite{thonhauser-prb06}.)  When
  viewing~(b) as the projected BZ of a layered 3D model in
  Sec.~\ref{sec:cycle-bulk}, the labels of the high symmetry points
  become $\gbar$, $\mbar$, $\kbar$, and $\kbar'$.}
\end{figure}

The Hamiltonian of a single layer $p$ is
\bea
\label{eq:H-monolayer}
H^{(0)}_p&=&(-1)^p\Delta\sum_{i}\tau_ic_{pi}^\dag c_{pi}
+t_1\sum_{\langle ij\rangle}c_{pi}^\dag c_{pj}\nn
&+&(-1)^p t_2\sum_{\la ij\ra}
\left(ic_{pi}^\dag c_{pj}+\text{h.c.}\right),
\eea
where indices $i$ and $j$ label sites on the $A$ and $B$ sublattices
marked in Fig.~\ref{fig:model}(a), and $\tau_i=+1$ ($-1$) for $i\in A$
($B$). The first and second terms contain the on-site energies and the
nearest-neighbor hoppings, respectively, and the third describes a
pattern of staggered magnetic fluxes generated by complex
second-neighbor hoppings; $\la ij\ra$ denotes pairs of sites for which
the hopping from $j$ to $i$ has amplitude $i(-1)^p t_2$, and ``h.c.''
stands for ``Hermitean conjugate.''  In the following we choose
$t_2>0$, and use it to set the energy scale for the model.

The Hamiltonian of the coupled bilayer system is
\beq
\label{eq:H-bilayer}
H_{\rm bilayer}=\sum_{p=1,2}H^{(0)}_p
+\tint\sum_i\tau_i\left(c_{1i}^\dag c_{2i}+\text{h.c.}\right),
\eeq
where $p=1$ and $p=2$ denote the layers at $z=0$ and $z=c/2$
respectively, and sites on different layers with the same index $i$
are aligned vertically. Thus, the first term describes two copies of
the Haldane model with opposite signs for both the on-site energies
and the complex hoppings, and the second couples them via vertical
hoppings that alternate in sign between the two sublattices.

We parameterize the model as
\begin{subequations}
\label{eq:cycle}
\begin{align}
 t_1&=-4t_2,\label{eq:t1}\\
 \tint&=t_2\label{eq:tint}\\
 \Delta&=
\left(3\sqrt{3} + 2\cos\cycle\right)t_2.\label{eq:Delta}
\end{align}
\end{subequations}
Referring to the phase diagram of the Haldane
model,\cite{haldane-prl88} the layer Chern numbers in the decoupled
limit $\tint\rightarrow 0$ are
\beq
C_p=
\begin{cases}
(-1)^{p-1}&\text{for }\cycle\in[\pi/2,3\pi/2]\text{ mod $2\pi$}\\
0&\text{otherwise}
\end{cases}.
\eeq
When $\cycle=\pi/2$~mod~$\pi$, the energy gap closes at~K on both
layers. The addition of interlayer hoppings reopens the gap, ensuring
that at half filling the system remains insulating for all values of
$\cycle$, with zero Chern number for the group of two valence bands.

\begin{figure}[tb]
\center
    \includegraphics[width=1.0\columnwidth]{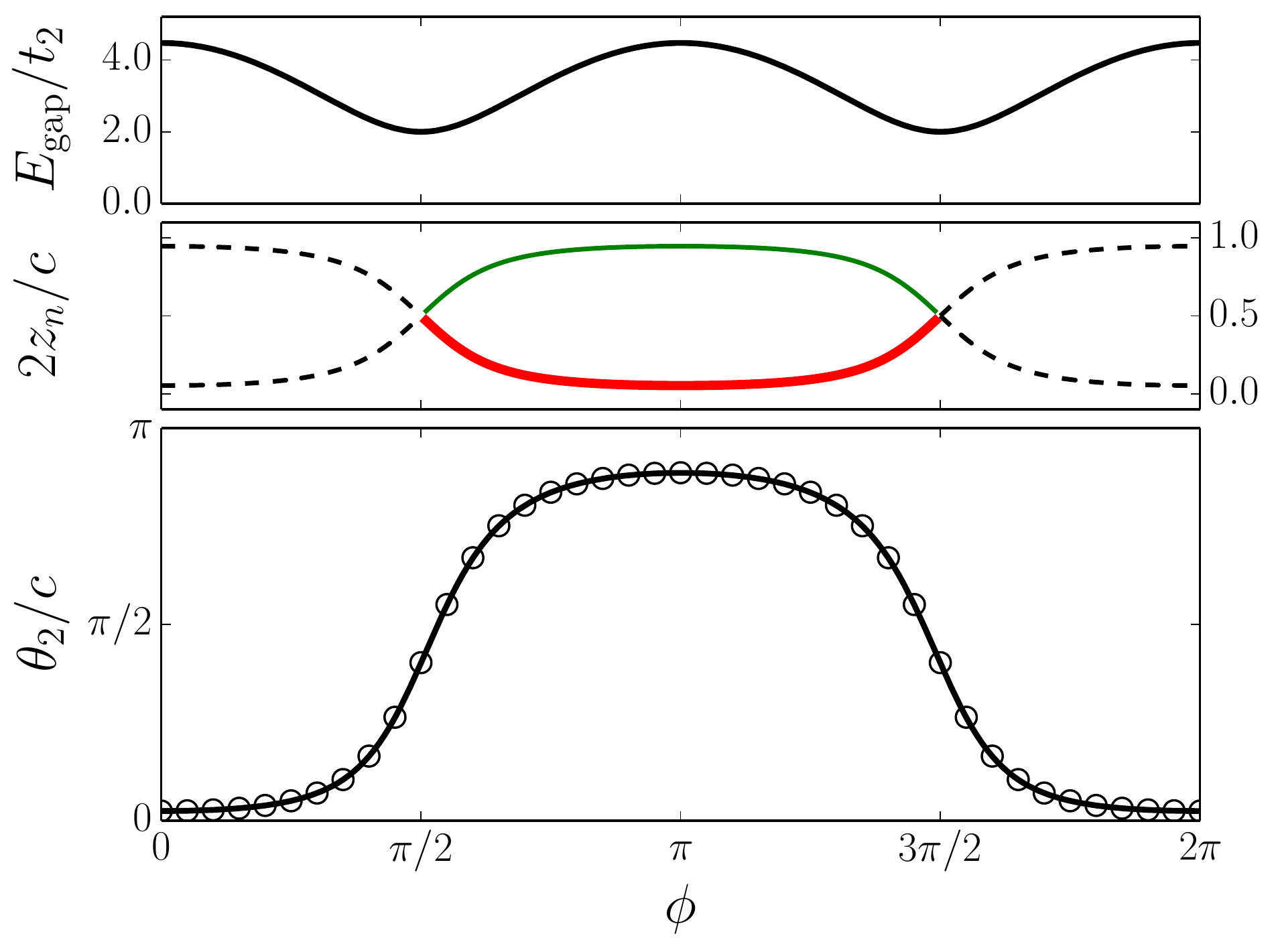}
    \caption{(Color online.) Numerical results for the bilayer model
      as a function of the adiabatic loop parameter $\cycle$.  Top:
      minimum energy gap between the second (highest occupied) and
      third bands -- the minimum gap is always at point~K in the 2D
      BZ. Middle: Wannier charge centers at~K, in units of the layer
      separation $c/2$. A dashed line indicates a Wannier sheet with
      Chern number $C=0$, a heavy (red) solid line denotes $C=+1$, and
      a light (green) solid line denotes $C=-1$.  The lower (upper)
      sheet has index $n=1$ ($n=2$). Bottom: dimensionless CSA
      coupling~$\theta_2/c$.  Open circles, extrapolation of
      $\theta_2(N)/c$ from finite-size samples. Solid line, direct
      calculation using the $k$-space
      formula~(\ref{eq:theta_2d_per}).}
\label{fig:bilayer}
\end{figure}

Although Chern numbers can no longer be defined for the individual
layers when $t_3\not= 0$, they are ``inherited'' by the Wannier
sheets.  The energy gap, Wannier centers at~K, and Wannier-sheet Chern
numbers are shown in the top and middle panels of
Fig.~\ref{fig:bilayer} over one cycle, $\cycle\in [0,2\pi]$.
Initially, the Chern numbers vanish for both sheets. The sheets
themselves are very flat, and sit close to one of the layers at $z=0$
or $z=c/2$.  With increasing~$\cycle$ they approach one another, at
$\cycle=\pi/2$ they touch at~K (but not at K$'$) changing their Chern
numbers to $\pm 1$, and then separate again.  There is no closure of
the energy gap during this process.  In the second half of the cycle
the system retraces the same parameter-space path in the opposite
direction.

The CSA coupling has been calculated in two different ways: for the
periodic 2D crystal using \eq{theta_2d_per} for $\theta_2$
(discretizing the covariant Berry curvature on a $120\times 120$
$k$-point mesh, following Ref.~\onlinecite{ceresoli-prb06}), and for
finite flakes using \eq{theta_0d} for $\theta_0$.  The results are
compared via \eq{0d-2d}, further dividing both sides by~$c$ (twice the
interlayer separation) in order to obtain a dimensionless coupling.
In practice we evaluate $\theta_2(N)=\tzero/N^2 A_{\rm cell}$, where
$\theta_0$ is for a flake containing $N\times N$ primitive cells, and
then extrapolate to $N\rightarrow\infty$ the results obtained for
$N=10,\,15,\,20,\,25,\,30$, by fitting to the
expression\cite{ceresoli-prb06}
\beq
\theta_2(N) = \theta_2(\infty)+a/N + b/N^2.
\eeq

The evolution of $\theta_2/c$ with $\cycle$ is shown in the bottom
panel of Fig.~\ref{fig:bilayer}.  It starts very small but nonzero,
and reaches a maximum of close to $\pi$ halfway through the cycle.
The precise agreement between the two curves confirms the correctness
of the $k$-space formula~(\ref{eq:theta_2d_per}). Importantly, the
agreement persists in the range $\pi/2<\phi< 3\pi/2$ where the Wannier
sheets have Chern numbers $\pm 1$.

\subsection{Cyclic evolution of the bulk model}
\label{sec:cycle-bulk}

The bulk model with two layers per cell and vertical lattice constant
$c$ is constructed by repeating the bilayer model along~$z$:
\beq
\label{eq:bulk}
 H_{\rm bulk}=\sum_p
\bigg[
H^{(0)}_p
+\tintl\sum_i\tau_i\left(c_{pi}^\dag c_{p+1,i}+\text{h.c.}\right)
\bigg]
\eeq
where the layer index $p$ now runs over all integers. The intralayer
hoppings and on-site energies are still given by \eqs{t1}{Delta}.  If
we were to use \eq{tint} for the interlayer hoppings, at half filling
the valence and conduction bands would touch at point~H in the 3D BZ
for $\cycle=\pi/2$ mod~$\pi$. In order to keep the gap open at all
$\cycle$, we modulate the interlayer hoppings out of phase with the
on-site energies,
\beq
\label{eq:t-int-bulk}
\tintl=
\left[
   1+(-1)^{p-1}\gamma\sin\cycle
\right]t_2
\eeq
with $\gamma=0.4$.  Each layer is more strongly coupled to the layer
below during half of the cycle, and to the layer above during the
other half. The insulating loop forms an ellipse on the
$(\Delta,\tintl)$ plane, with a metallic point at the center.

\begin{figure}[tb]
\center
    \includegraphics[width=1.0\columnwidth]{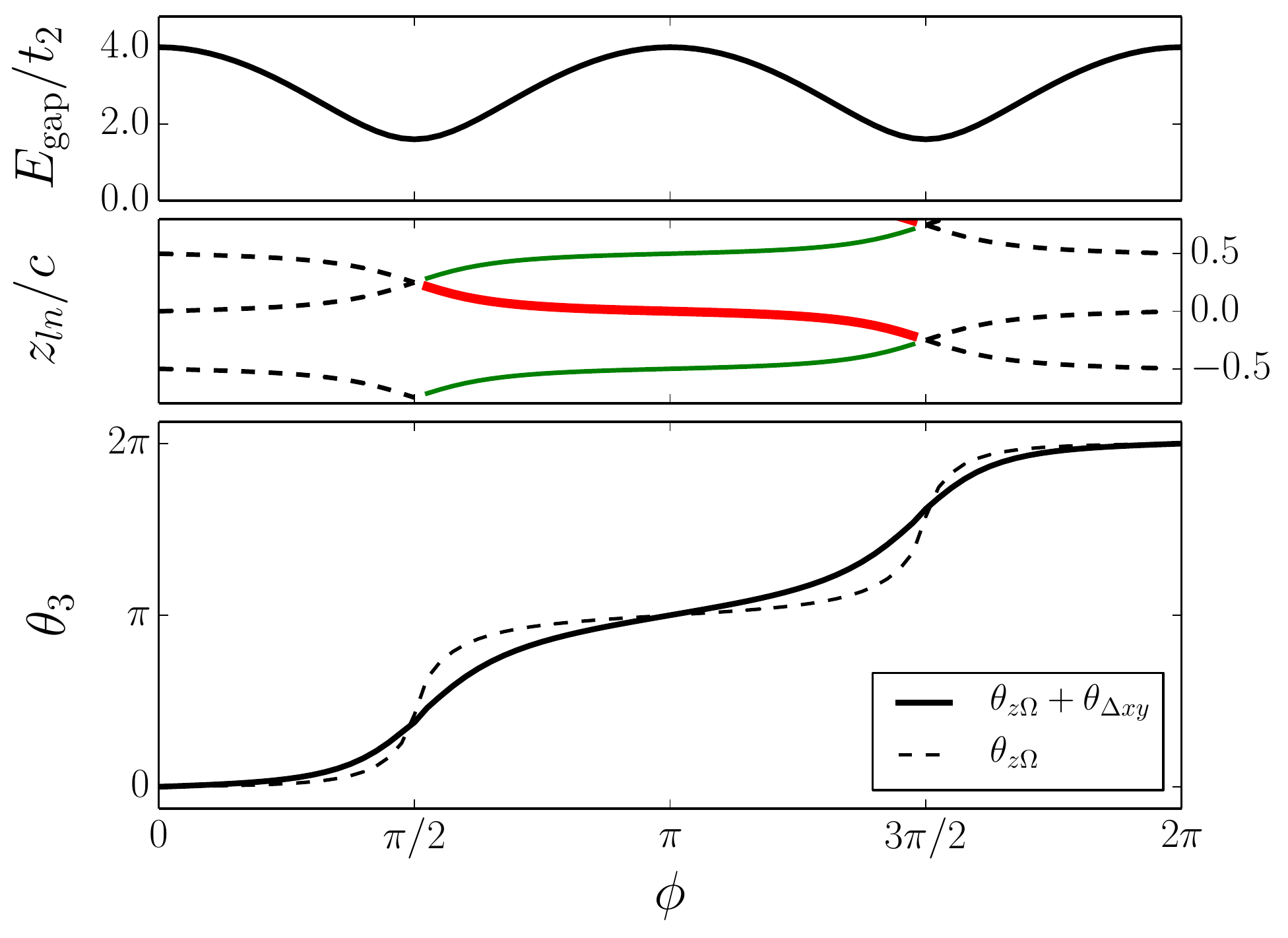}
    \caption{(Color online.) Numerical results for the bulk model as a
      function of the adiabatic loop parameter $\cycle$. Top: minimum
      energy gap between the second and third bands -- the minimum gap
      is always at point~H, which projects onto $\kbar$ in the surface
      BZ.  Middle: periodically-repeated Wannier charge centers
      at~$\kbar$, in units of the lattice constant $c$; the dashed
      line and colored heavy/light lines have the same meaning as in
      Fig.~\ref{fig:bilayer}. The lower Wannier sheet has indices
      $l=-1$ and $n=2$, the middle one $l=0$ and $n=1$, and the upper
      one $l=0$ and $n=2$. Bottom: total CSA
      coupling~$\tthree=\theta_{z\Omega}+\theta_{\Delta xy}$, and the
      term $\theta_{z\Omega}$ responsible for pumping.}
\label{fig:bulk}
\end{figure}

To demonstrate the pumping behavior of this model we have calculated
$\tthree(\cycle)$ from \eq{theta_3d_tot}.
The HWFs and the charge centers were obtained from a
parallel-transport construction applied to strings of $k$~points along
$k_z$.\cite{marzari-prb97}

The results obtained with a mesh of $120\times 120\times 6$ points in
the BZ are shown in Fig.~\ref{fig:bulk}. The energy gap remains open
throughout the cycle. In the first half, the evolution of the two
Wannier centers in the home cell $l=0$ resembles that in
Fig.~\ref{fig:bilayer} for the bilayer system. At point $\kbar$ in the
projected BZ they start off at $0$ and $c/2$ respectively, begin
approaching until at $\cycle=\pi/2$ they touch at $z=c/4$ exchanging
one unit of Chern number, and then drift apart, recovering the initial
separation at $\cycle=\pi$.  By then $\tthree$ has increased
continuously from $0$ to $\pi$, with the term $\theta_{\Delta xy}$
going through zero at $\cycle=0,\,\pi/2$, and ~$\pi$.

Thanks to the modulation in the interlayer hoppings, the second half
of the cycle is very different from that of the bilayer. Instead of
reconnecting with their original partners, the Wannier centers
continue to drift in the same direction until at $\cycle=3\pi/2$ they
touch the periodic images of their original partners in the adjacent
cells.  As a result $\tthree$ continues to increase, reaching $2\pi$
by the end of the cycle.

At $\cycle=0 \text{ mod $\pi$}$ the model acquires extra
symmetry:\footnote{In addition to mirror symmetry $M_z$, the bulk
  model at $\phi=0,\pi$ also exhibits $C_2^y*\trev$ symmetry, with the
  two-fold axis lying on the plane halfway between two atomic layers
  and pointing along a zig-zag edge. This additional symmetry, which
  also constrains $\theta_3$ to be an integer multiple of $\pi$, can
  be removed by changing the neareast-neighbor hopping amplitude along
  one of the two in-plane directions. We have checked that doing so
  modifies the $\theta_3(\phi)$ curve, but does not affect the
  quantized values at $\phi=0,\pi$.}
each layer becomes a mirror plane, forcing $\tthree$ to be an integer
multiple of $\pi$.\cite{fang-prb12,varjas-prb15} According to
Fig.~\ref{fig:bulk}, $\theta_3=0$ at $\phi=0$ and $\theta_3=\pi$ at
$\phi=\pi$.  In the CSA pump model of
Ref.~\onlinecite{taherinejad-prl15}, the same values occured at the
beginning and in the middle of the pumping cycle, where they were
protected by~$\trev$ rather than mirror symmetry.  In
Appendix~\ref{app:mirror} we analyze further the quantization of
$\theta_3$ due to mirror symmetry in our model, and show that at
$\cycle=\pi$ it becomes a topological crystalline insulator with a
single surface Dirac cone on any surface normal to the mirror plane.

The fact that the CSA coupling is pumped by $2\pi$ over an adiabatic
cycle signals that the occupied band manifold of the 4D Bloch
Hamiltonian $H_{\rm bulk}(k_x, k_y, k_z, \cycle)$ has a nonvanishing
second Chern number.\cite{qi2008} In Appendix~\ref{app:2-param} we map
the second Chern number of the model over an augmented parameter
space.

\subsection{Cyclic evolution of the slab interior, keeping the
  surfaces gapped}

Let us now study the cyclic evolution in a slab geometry.  If the
entire slab --~including the surfaces~-- returns to the initial state
at the end of the cycle, it must pass through a state with metallic
surfaces.~\cite{taherinejad-prl15} This scenario will be investigated
in the next section, but first we examine here what happens when a
surface modification is introduced to avoid the gap closure at the
surfaces. To this end we adjust the on-site energies on the top and
bottom layers of the slab according to
\beq
\label{eq:surface_mod}
\Delta=\left(3\sqrt{3}-2\right)t_2\,\,\,\,\,\text{for $\pi<\cycle\leq 2 \pi$,}
\eeq
so that in the second half of the cycle the on-site energies on the
two surface layers remain frozen at the values reached at
$\cycle=\pi$, while on all other layers they continue to evolve
according to \eq{Delta}.  Thus, at $\phi=2\pi$ the surfaces are in a
{\it different} insulating state than at $\phi=0$. According to the
surface theorem the CSA couplings of the two configurations may differ
by a multiple of $2\pi$, and this is indeed what happens for our
model.

\begin{figure}[tb]
\center
    \includegraphics[width=1.0\columnwidth]{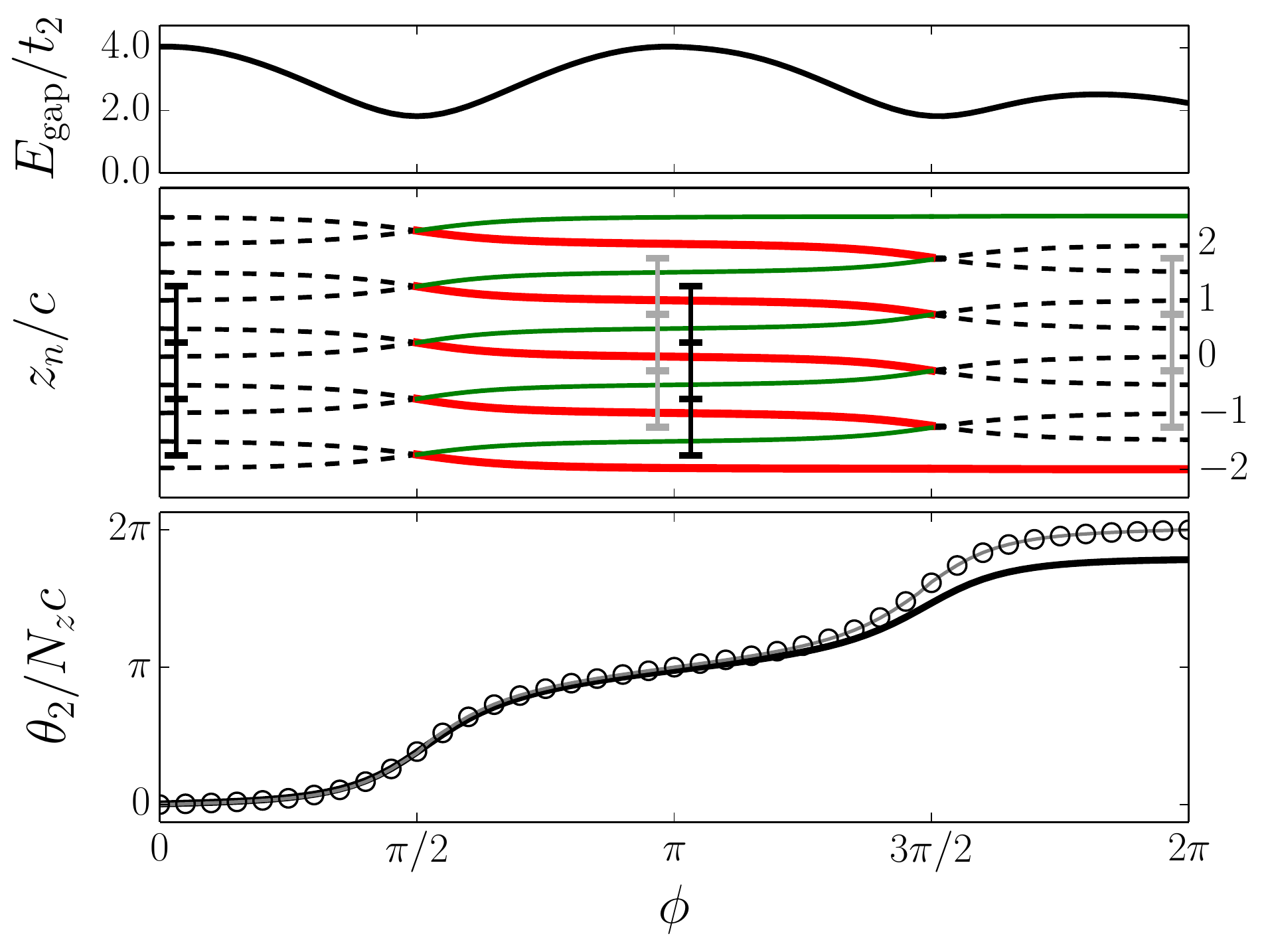}
    \caption{(Color online.) Numerical results for a ten-layer slab
      ($N_z=5$) as a function of the adiabatic loop parameter
      $\cycle$. The surface Hamiltonian has been modified according to
      \eq{surface_mod}, to keep the surface insulating throughout the
      cycle.  Top: minimum energy gap (always at point~K in the 2D BZ)
      at half filling.  Middle: Wannier charge centers at~K in units
      of the lattice constant~$c$, with indices ranging from $n=1$
      (bottom) to $n=10$ (top); the dashed line and colored
      heavy/light lines have the same meaning as in
      Fig.~\ref{fig:bilayer}. Two possible choices of bulk cells are
      indicated by the grey and black lines.  Bottom: dimensionless
      CSA coupling $\theta_{\rm slab}=\theta_2/N_zc$.  Heavy line,
      calculation done at $N_z=5$.  Open circles, extrapolation to
      $N_z\rightarrow\infty$ of calculations done at $N_z=4,\,7,10$.
      Light grey line, bulk CSA coupling $\tthree$ taken from
      Fig.~\ref{fig:bulk}.}
\label{fig:slab-pump}
\end{figure}

We consider a slab containing $N_z$ repetitions in the~$z$ direction
of the model defined by \eq{bulk}, and use \eq{theta_2d_per} to
calculate $\theta_{\rm slab}(\phi)$ [\eq{theta-slab}] at half filling.
Figure~\ref{fig:slab-pump} shows the results obtained with a ten-layer
slab, i.e., $N_z=5$. Thanks to the surface modification, the energy
gap shown in the top panel remains open throughout the cycle.  As seen
in the middle panel, all ten Wannier sheets start out with zero Chern
number. The eight subsurface Wannier sheets show a bulklike behavior,
switching partners between $\cycle=\pi/2$ and $\cycle=3\pi/2$ as in
Fig.~\ref{fig:bulk}. This leaves the top and bottom sheets unpaired
and with Chern numbers $-1$ and $+1$ respectively until the end of the
cycle.  The slab CSA coupling (heavy line in the lower panel) goes
from zero at $\phi=0$ to somewhat less than $2\pi$ at $\phi=2\pi$.  It
is only when $N_z\rightarrow \infty$ (open circles) that
$\theta_{\rm slab}(2\pi)\rightarrow 2\pi$; this is similar to an
adiabatic charge pump, where the exact quantization of particle
transport only occurs in the thermodynamic limit.\cite{thouless-prb83}

The result $\theta_{\rm slab}(2\pi)-\theta_{\rm slab}(0)=2\pi$ in the
limit of a thick slab can be readily understood from the surface
theorem.  Since $H_{\rm bulk}(\phi)$ is periodic, the sheet structure
in the interior of the slab is identical at $\cycle=0$ and at
$\cycle=2\pi$, so that $\tthree(0)=\tthree(2\pi)$ for any choice of
bulk cell (two possible choices are indicated in
Fig.~\ref{fig:slab-pump}; since all the bulk sheets have zero Chern
number, the value of $\theta_3$ does not depend on that
choice\cite{taherinejad-prl15}).  What changes between the two states
of the slab is the excess Chern number that remains after tiling the
chosen cell towards the top surface: inspection of
Fig.~\ref{fig:slab-pump} shows that $C_{\rm III}(0)=0$, while
$C_{\rm III}(2\pi)=-1$.  Referring to \eq{surf-thm},
\beq
\left.\Delta\theta_{\rm slab}\right|_{\cycle=0}^{\cycle=2\pi}
=-2\pi\left.\Delta C_A\right|_{\cycle=0}^{\cycle=2\pi}=2\pi.
\eeq

Let us now use \eq{surf-thm} to determine $\theta_{\rm slab}$ in the
range $\pi/2<\cycle < 3\pi/2$ where the Wannier sheets have
alternating Cherns numbers $\pm1$.  When switching from the grey to
the black cells in Fig.~\ref{fig:slab-pump}, the sheet with $C=+1$
(heavy red line) in the central cell stays the same, while the one
with $C=-1$ (light green line) gets replaced by its periodic image
below, changing $\tthree$ by $-2\pi$. Tiling the former cell towards
the upper surface leaves no excess sheets, so that $C_{\rm III}=0$ and
$\theta_{\rm slab}=\tthree$.  With the latter choice there is one
leftover sheet with $C_{\rm III}=-1$ that contributes $+2\pi$ to
$\theta_{\rm slab}$, exactly cancelling the change in $\tthree$. The
calculated value of $\theta_{\rm slab}$ is therefore the same with
both cell choices.

\subsection{Cyclic evolution of the entire slab}

\begin{figure}[tb]
\center
    \includegraphics[width=0.98\columnwidth]{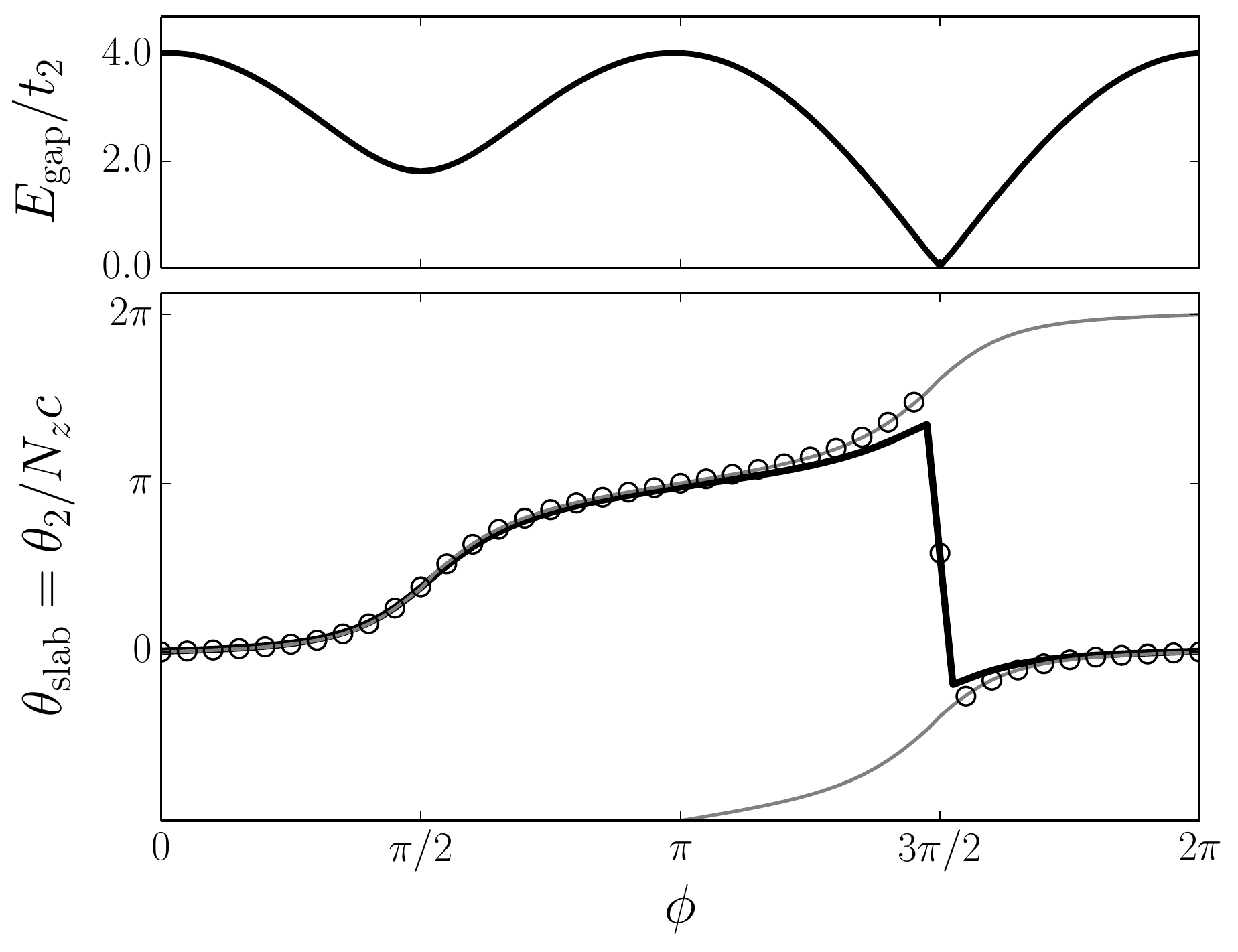}
    \caption{Same as the top and bottom panels of
      Fig.~\ref{fig:slab-pump}, except that now the entire slab
      --~including the surfaces~-- undergoes a cyclic evolution.  }
\label{fig:slab-no-pump}
\end{figure}

Next we consider a truly cyclic evolution of the slab, where not only
the interior but also the surfaces return to their initial state.
Because $\theta_{\rm slab}(\cycle)$ is uniquely determined by
$H_{\rm slab}(\phi)$ and the band filling, it must return at
$\phi=2\pi$ to the same value it had at $\phi=0$.  The way this
happens for our model is shown in Fig.~\ref{fig:slab-no-pump}. From
$\phi=0$ to $\phi_c=3\pi/2$ the CSA coupling of a thick slab (open
circles) evolves exactly as in Fig.~\ref{fig:slab-pump}. Then at the
critical parameter $\phi_c$ it drops abruptly by $2\pi$, switching
between two branches of the bulk CSA coupling $\tthree$ (light grey
lines).

\begin{figure}[tb]
\center
    \includegraphics[width=1.0\columnwidth]{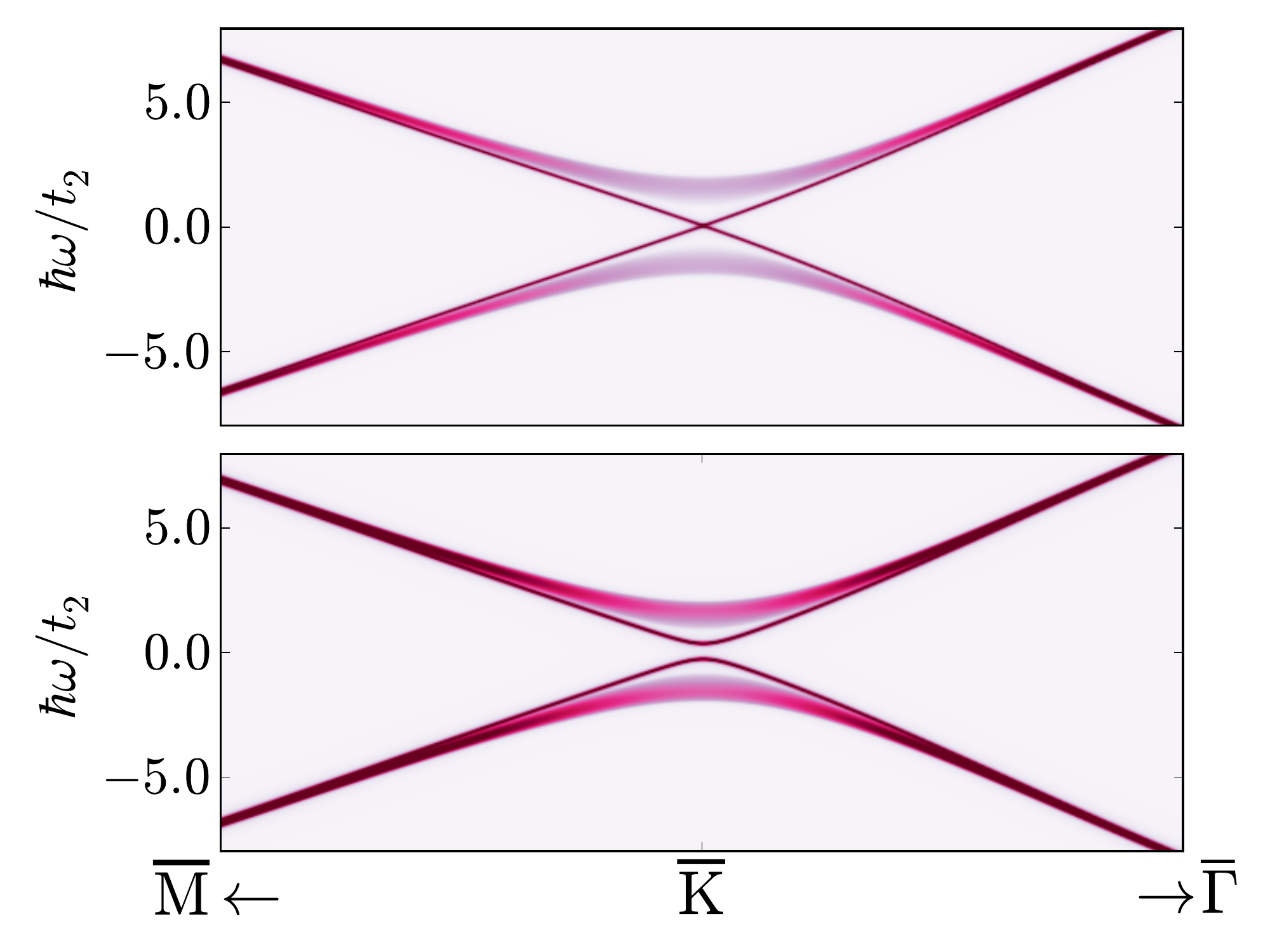}
    \caption{(Color online.) Surface spectral function in a
      semi-infinite geometry, calculated at the critical parameter
      value $\phi_c=3\pi/2$ where the CSA coupling in
      Fig.~\ref{fig:slab-no-pump} changes discontinuously (top panel),
      and at $\phi=\phi_c\pm \pi/20$ (bottom panel).  The spectral
      function is plotted along two lines that extend $(1/20)$th of
      the distance from $\kbar$ to $\mbar$ and $\gbar$.}
\label{fig:weyl}
\end{figure}

At $\cycle_c$, the energy gap closes at the two surfaces. In order to
avoid spurious interactions between the surfaces that reopen the gap
slightly and complicate the analysis, let us switch to a semi-infinite
geometry with a single surface.  Figure~\ref{fig:weyl} shows the
spectral function in the vicinity of point $\kbar$ in the surface BZ,
evaluated by tracing over sites in the topmost two layers. The gap
closing at $\cycle_c$ (top panel) consists of a linear crossing
between two surface bands.  Moving sligthly away from $\phi_c$ (bottom
panel), the crossing becomes avoided. Just like the touching events
between Wannier sheets seen in
Figs.~\ref{fig:bilayer}--\ref{fig:slab-pump}, this band touching
constitutes a Weyl point in the space of parameters
$(k_x,k_y,\phi)$. Since the states that cross are localized at the
surface, we call it a ``surface Weyl point.''

The discontinuous $2\pi$ drop in $\theta_{\rm slab}$ occurs because at
$\phi_c$ a $-2\pi$ quantum of Berry flux is channeled from the valence
to the conduction bands through the surface Weyl point. This removes
the exact amount of Berry curvature that gets pushed to the surface
during one pumping cycle, allowing the surface AHC to return to its
initial value by the end of the cycle. Referring to the middle panel
of Fig.~\ref{fig:slab-pump} (but without the bottom surface, due to
the semi-infinite geometry), the effect of the surface Weyl point
would be to add at $\phi_c=3\pi/2$ a Chern amount of $+1$ to the
topmost Wannier sheet, changing its Chern number from $-1$ to zero
without it touching the sheet below.

The scenario sketched here, in which a single Weyl point in
$(k_x,k_y,\phi)$ space disposes of the excess Chern number pumped to
the surface during a cycle that also returns the surface Hamiltonian
to itself, is not the only possibility.  In general there could be a
finite interval in $\cycle$ over which the surface becomes metallic.
In such cases, however, we would still expect that the ``Fermi surface
in $(k_x,k_y,\phi)$ space'' should always enclose such a Weyl
point,\cite{gosalbez-prb15} or more precisely, a set of Weyl points
with a net chirality equal to the second Chern number characterizing
the pumping cycle.

\section{Summary}

In summary, we have examined how the presence of insulating surfaces
fixes the quantized part of the CSA coupling that is undefined for a
purely bulk band insulator.  In the basis of HWFs maximally-localized
along the surface-normal direction, the CSA coupling of a thick slab
becomes a sum of two terms: (i) a nonquantized contribution
(previously found in Ref.~\onlinecite{taherinejad-prl15}) associated
with the bulklike HWFs far from the surfaces, and (ii) a quantized
contribution given by the excess Chern number of the Wannier sheets
near the surface.  When some of the bulk Wannier sheets have nonzero
Chern numbers the individual terms in this decomposition become
dependent on the choice of bulk cell, but their sum remains unique
(for a given surface termination).

Inspired by the representation of the CSA pumping process in the HWF
basis, where $2\pi$ quanta of Berry curvature are passed from sheet to
sheet,\cite{taherinejad-prl15} we constructed a 3D tight-binding model
for a quantum CSA pump by coupling quantum anomalous Hall layers with
tunable Chern numbers.  In order to illustrate the surface theorem of
\eq{surf-thm}, the CSA coupling strength of the model, as well as the
charge centers and Chern numbers of the individual Wannier sheets,
were tracked during cyclic evolutions carried out in different
geometries (periodic bulk crystal, finite slab, and semi-infinite
crystal). These numerical studies revealed how the Berry curvature
pumped across the bulk is extracted at the surfaces, when the surface
Hamiltonian also undergoes a cyclic evolution.

At two isolated points along the cycle, our model acquires mirror
symmetry. The two mirror-symmetric states are topologically distinct:
it is not possible to go from one to the other along a path in
parameter space that preserves mirror symmetry without closing and
reopening the direct band gap.  The trivial state has $\theta_3=0$~mod
$2\pi$ (corresponding to a quantized surface AHC of $ne^2/h$), while
the nontrivial one has $\theta_3=\pi$~mod $2\pi$ and a {\it
  half}-quantized surface AHC, and these are the only values
consistent with mirror symmetry.  This is analogous to the
quantization of the bulk polarization and ``surface'' (edge) bound
charge to integer and half-integer multiples of $e$, respectively, at
the two inversion-symmetric points along a charge-pumping cycle in
1D.\cite{vanderbilt-prb93}

The HWF representation has become a powerful tool for identifying
topological phases in 2D and
3D.\cite{fu-prb06,soluyanov-prb11,yu-prb11,taherinejad-prb14,gresch-arxiv16}
While in previous studies the objects of interest were the bulk HWFs,
the present work shows that, as in the 1D polarization problem, in 3D
additional topological information can be extracted from the HWFs near
the surface.

\begin{acknowledgments}

  The Center for Nanostructured Graphene (CNG) is sponsored by the
  Danish National Research Foundation, Project DNRF103.
  D.V. acknowledges support from NSF Grant DMR-1408838.
  I.S. acknowledges support from the European Commission Grant
  No. CIG-303602, and from grant No.~FIS2016-77188 from the Spanish
  Ministerio de Econom\'ia y Competitividad.

\end{acknowledgments}

\appendix

\section{Quantization of the CSA coupling in the presence of mirror
  symmetry}
\label{app:mirror}

In this Appendix we use the HWF picture to analyze how the presence of
mirror symmetry $M_z$ in our bulk model at $\phi=0$ and $\phi=\pi$
forces the CSA coupling to be either 0 or $\pi$ modulo $2\pi$, as seen
in Fig.~\ref{fig:bulk}. Henceforth we use the symbol $\theta^M_3$ to
denote the value of the CSA coupling when mirror symmetry is present.

Mirror symmetry forces the term $\theta_{\Delta xy}$ in \eq{dxy} to
vanish, because that term is odd under $M_z$ and single valued. Therefore,
$\theta_3^M$ is fully determined by \eq{zomega},
\beq 
\label{eq:theta-M}
\theta_3^M=-\frac{1}{c}\int d\kpar\sum_{n}
  z_{\kpar 0n}\Omega^{xy}_{\kpar,0n,0n}.
\eeq

According to Fig.~\ref{fig:bulk}, at $\cycle=0$ and $\cycle=\pi$ the
two Wannier centers at $\kbar$ in the home cell of our model coincide
with one of the two mirror planes, $z=0$ or $z=1/2$ (in units of the
lattice constant~$c$).  In fact, the two Wannier sheets are pinned to
the mirror planes for all values of $\kpar$. Taking the Wannier
centers out of the integrand in the previous equation and using
\eq{chern}, we find
\beq
\theta^M_3=-\pi C_{\nicefrac{1}{2}},
\eeq
where $C_{\nicefrac{1}{2}}$ is the Chern number of the Wannier sheet
pinned at $z=1/2$. Consulting again the middle panel of
Fig.~\ref{fig:bulk} we see that $C_{\nicefrac{1}{2}}=0$ at $\phi=0$
and $C_{\nicefrac{1}{2}}=-1$ at $\phi=\pi$, so that $\theta^M_3=0$ and
$\theta^M_3=\pi$ respectively, in agreement with the lower panel of
the same figure.

Alternatively, the band topology protected by mirror symmetry can be
characterized by ``mirror Chern numbers'' $C^M_{k_z}$ defined on the
mirror-invariant planes in the BZ\cite{teo-prb08} (for this model,
they are $k_z=0$ and $k_z=\pi$, in units of~$1/c$).  The two
descriptions are related by\cite{varjas-prb15}
\beq
\label{eq:theta-mirror}
\theta_3^M
=\pi\left(\left|C^M_0\right|
+\left|C^M_\pi\right|\right)\,\text{mod $2\pi$}.
\eeq
Below we verify this relation explicitly for our model. We make use of
the fact that $\left| C^M_{k_z}\right|$ counts the number of Dirac
cones along the projection line of the $k_z$ plane onto the surface BZ
of a mirror-symmetric surface.\cite{teo-prb08}

\begin{figure}[tb]
\center
    \includegraphics[width=1.0\columnwidth]{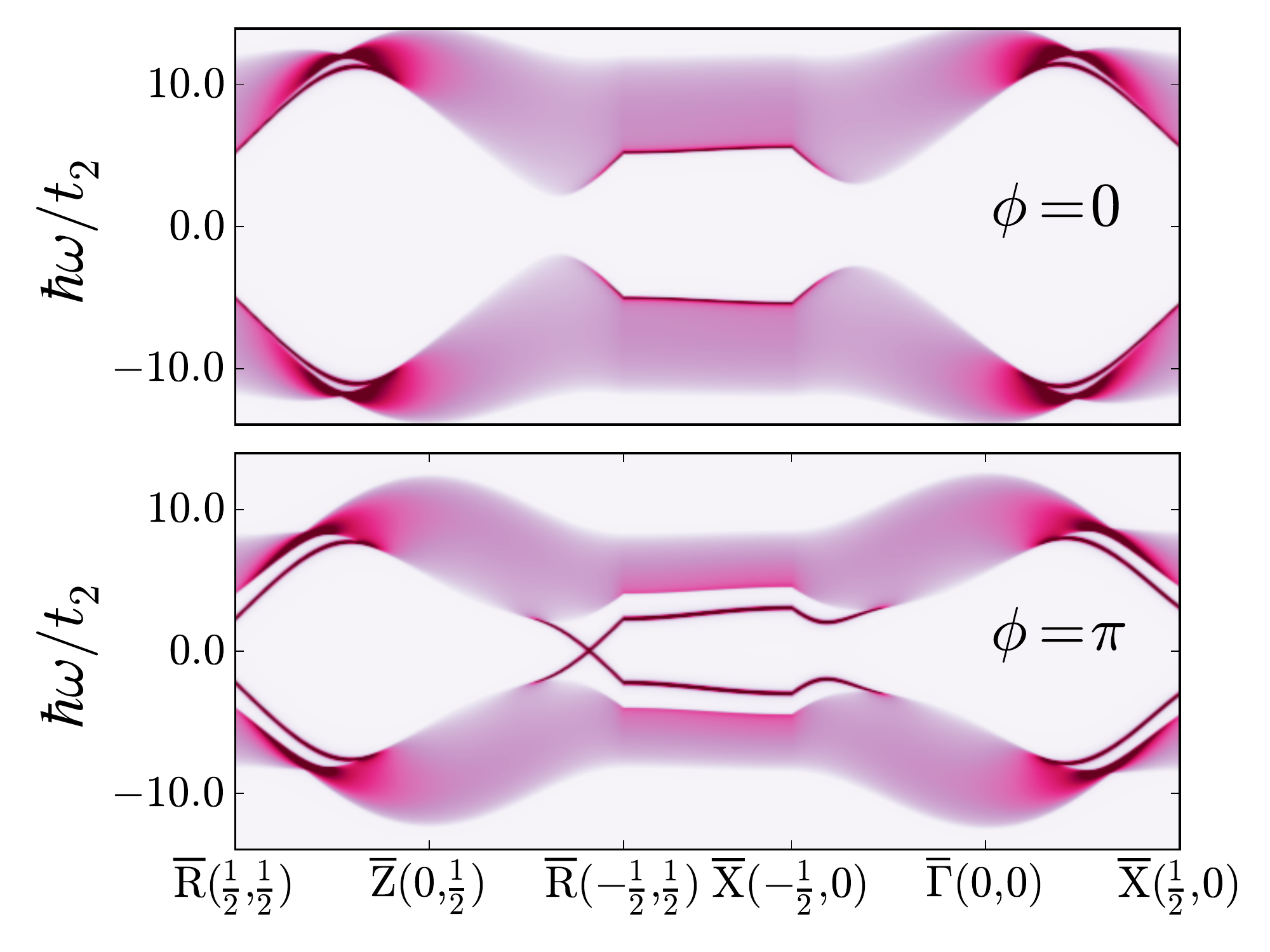}
    \caption{(Color online.) Surface spectral function of the layered
      model at $\cycle=0$ and $\cycle=\pi$, for a semi-infinite
      geometry with the surface orthogonal to the $y$~direction.  In
      both cases the entire system (bulk plus surface) has mirror
      symmetry, but while in the top panel the bulk CSA coupling
      vanishes, in the bottom panel it equals~$\pi$ and protects a
      surface Dirac cone.}
\label{fig:mirror}
\end{figure}

Figure~\ref{fig:mirror} shows the surface spectral function calculated
at $\cycle=0,\pi$ using a semi-infinite geometry that respects mirror
symmetry (surface normal to the $y$~direction, corresponding to zig
zag edges for the individual layers). The surface BZ is a rectangle on
the $(k_x, k_z)$ plane, with $\overline{\Gamma}=(0,0)$ at the center,
$\overline{\rm R}=(\nicefrac{1}{2},\nicefrac{1}{2})$ at the corner,
and $\overline{\rm X}=(\nicefrac{1}{2},0)$ and $\overline{\rm
  Z}=(0,\nicefrac{1}{2})$ at the centers of the two edges.

At $\cycle=0$ (top panel of Fig.~\ref{fig:mirror}), the absence of
surface states traversing the bulk gap implies
$\left| C^M_0\right|=\left| C^M_\pi\right|=0$, in agreement with the
value $\theta_3=0$.  At $\cycle=\pi$ (bottom panel) there are two such
surface bands of opposite chirality.  They cross along the projection
$\overline{\rm Z}\,\overline{\rm R}$ of the $k_z=\pi$ plane so that
$\left| C^M_\pi\right|=1$ and $\left| C^M_0\right|=0$, consistent with
$\theta_3=\pi$.  In conclusion, the model at $\cycle=0$ is a trivial
insulator while at $\cycle=\pi$ it is a topological crystalline
insulator with a single Dirac cone on any surface normal to the mirror
plane.

\section{Mapping the second Chern number in an augmented parameter
  space }
\label{app:2-param}

In the same way that a quantum charge pump is characterized by a
nonzero (first) Chern number defined over a two-dimensional parameter
space $(k,\lambda)$,\cite{thouless-prb83} a quantum CSA pump has a
nonvanishing second Chern number in the four-dimensional space of
$\bk$ and the pumping parameter,\cite{qi2008} i.e., $(k_x, k_y, k_z,
\cycle)$ for the model of \eq{bulk}.

\begin{figure*}[tb]
\center
    \includegraphics[width=5.8 cm]{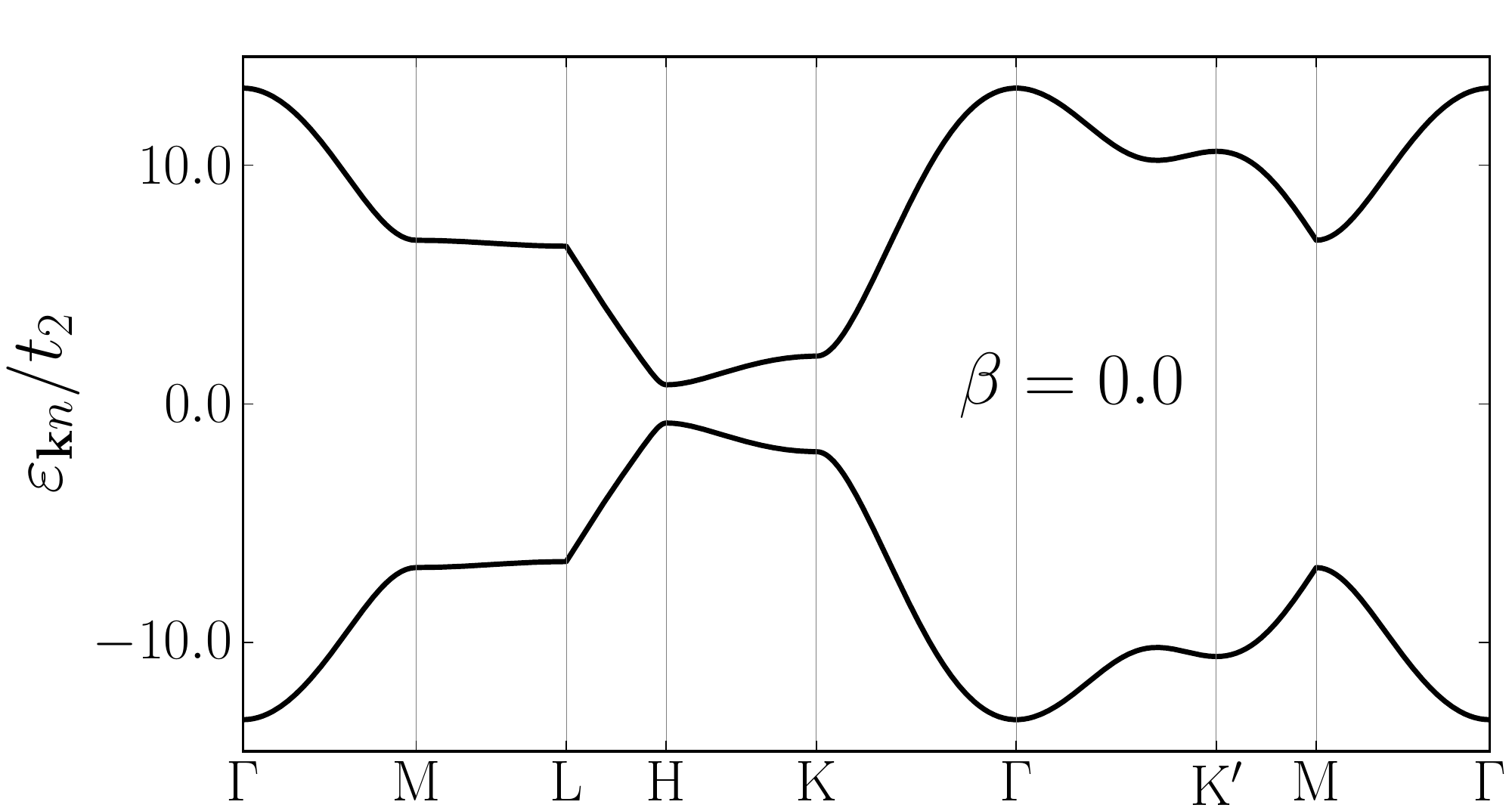}
    \includegraphics[width=5.8 cm]{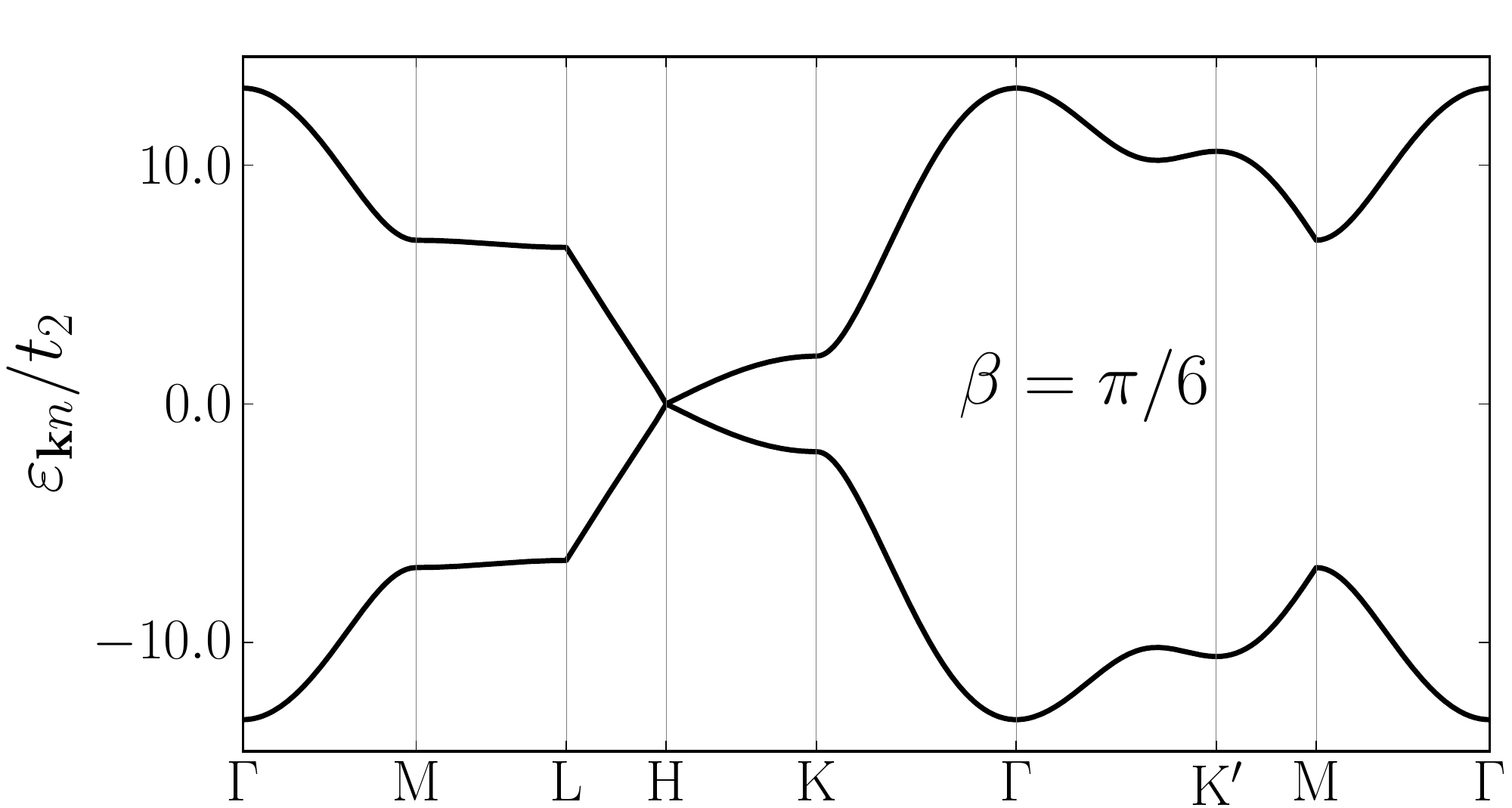}
    \includegraphics[width=5.8 cm]{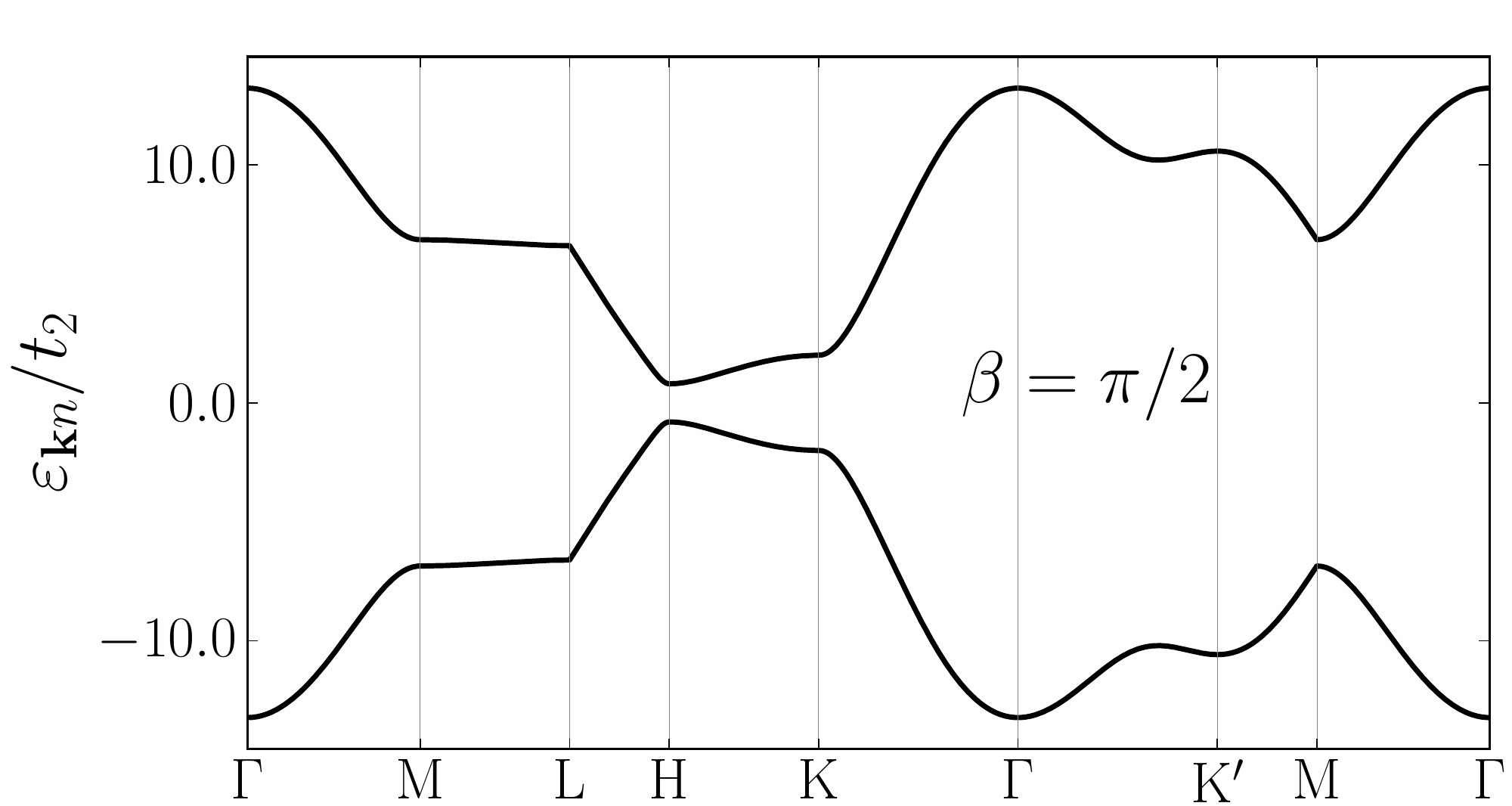}
    \includegraphics[width=5.8 cm]{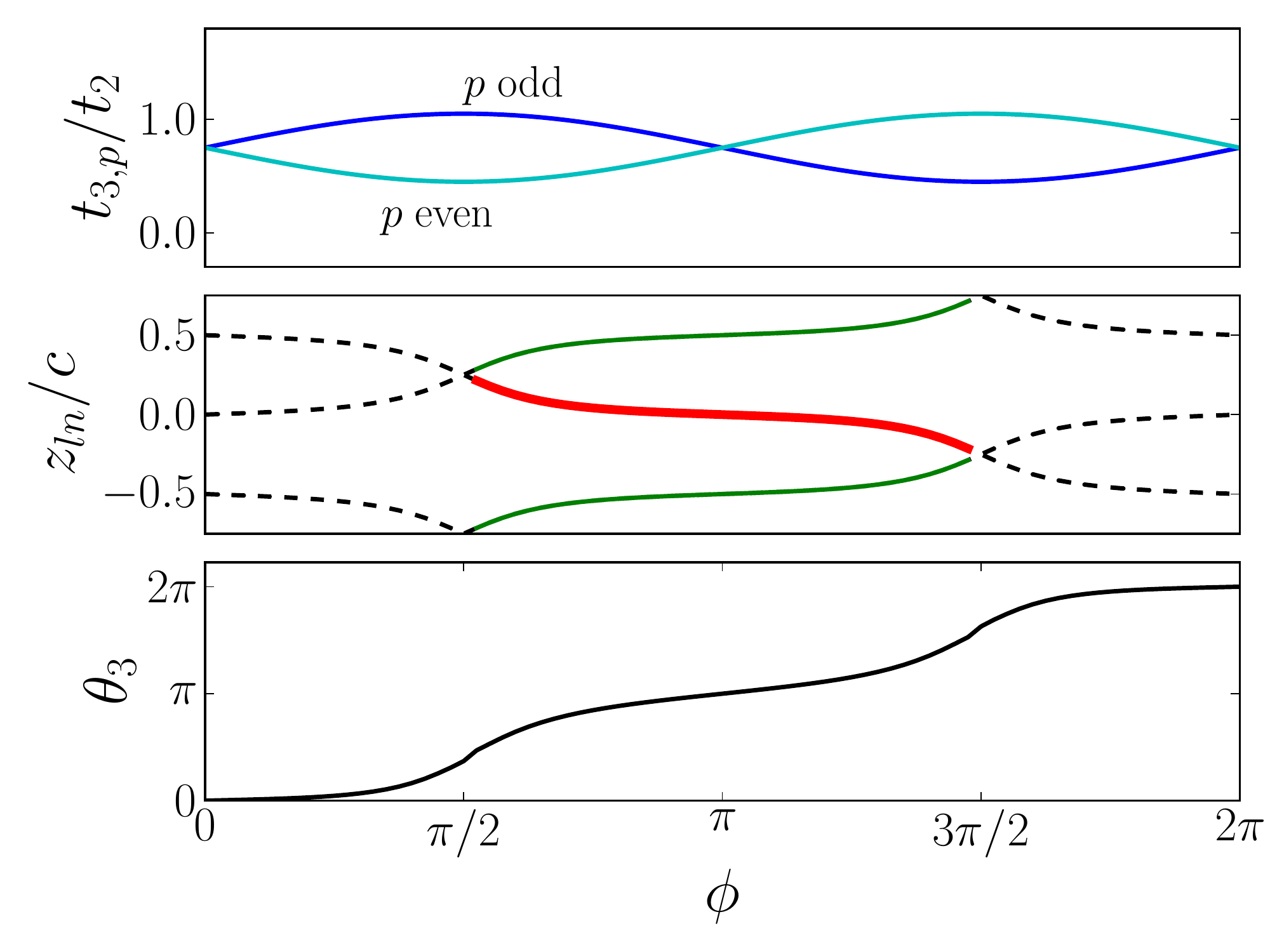}
    \includegraphics[width=5.8 cm]{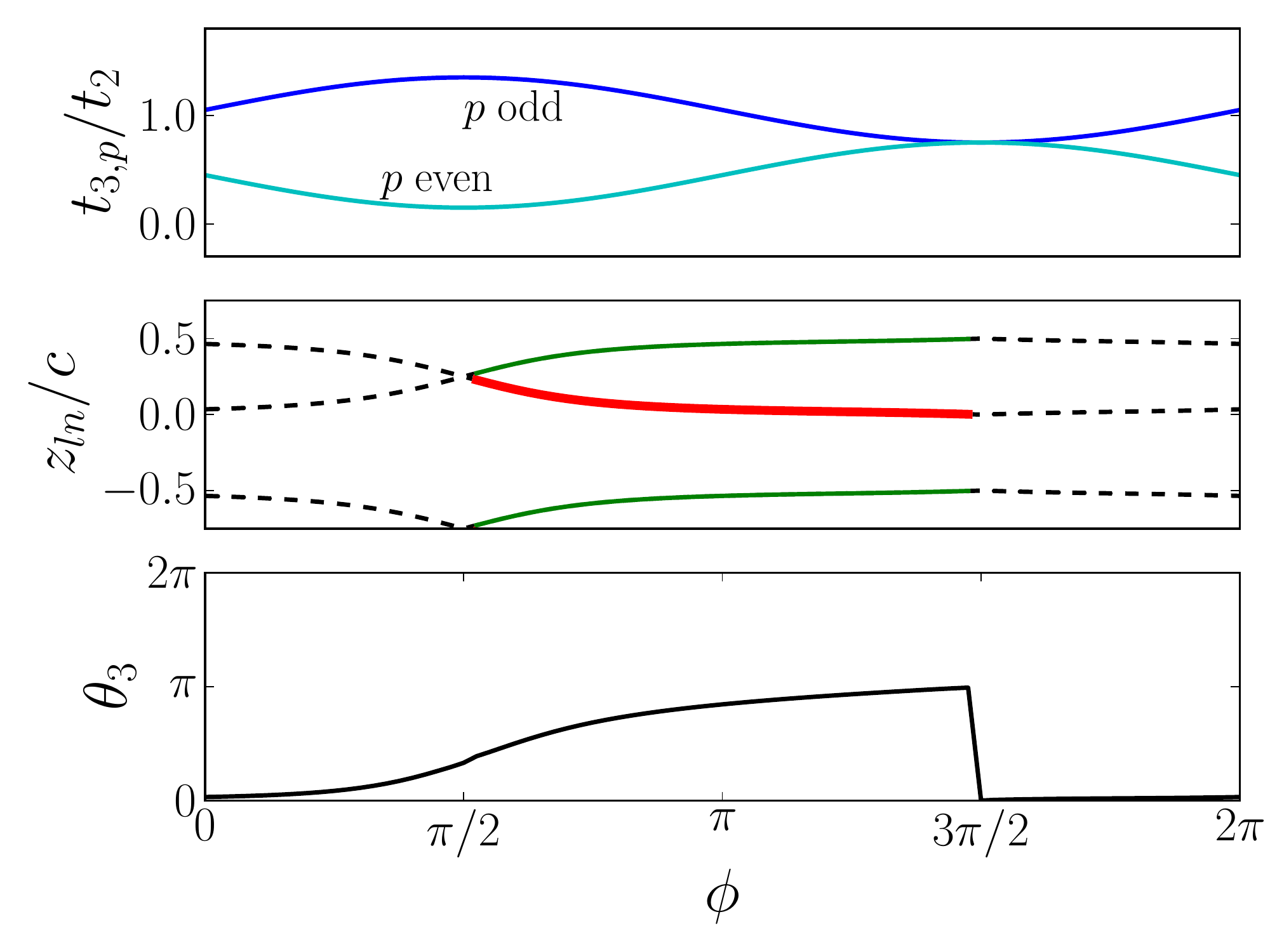}
    \includegraphics[width=5.8 cm]{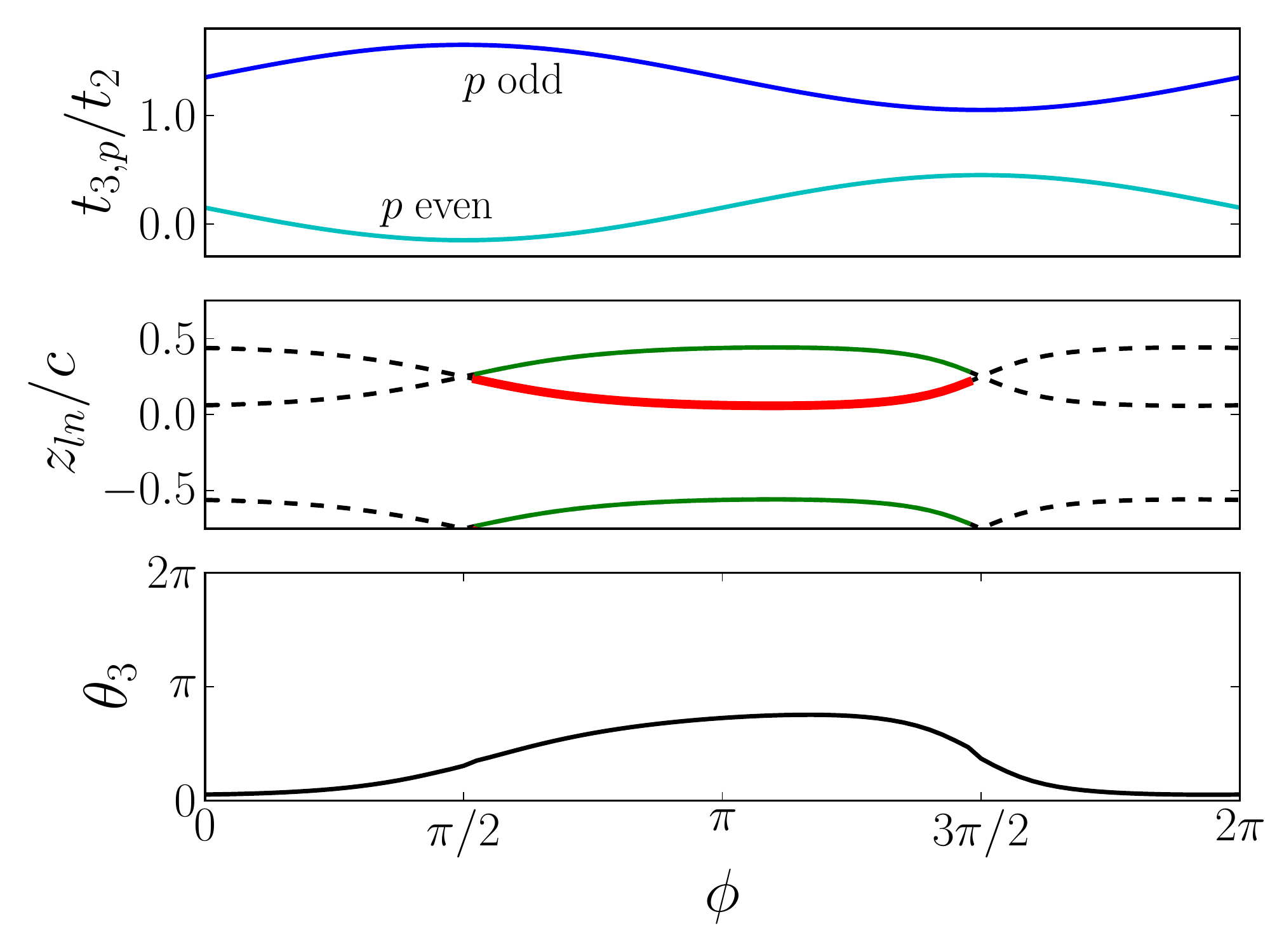}    
    \caption{(Color online.) Top: band structures in the 3D BZ,
      calculated at $\cycle=3\pi/2$ with three different values of
      $\beta$. The bands are everywhere two-fold degenerate, and at
      $\beta=\pi/6$ they become four-fold degenerate at~H, forming a
      Dirac point. Bottom: interlayer couplings, hybrid Wannier charge
      centers at point $\kbar$ in the projected BZ, and bulk CSA
      coupling, plotted as a function of $\phi$ for the same three
      values of~$\beta$. In the panels depicting the Wannier centers,
      the dashed lines and colored heavy/light lines have the same
      meaning as in Fig.~\ref{fig:bilayer}.}
\label{fig:dirac}
\end{figure*}

In order to vary the behavior of the loop so that it can also
represent trivial cycles, we introduce an additional parameter $\beta$
in the interlayer hoppings of the model,
\begin{align}
\label{eq:t-int-2param}
\tintl&= \left[ 1 +(-1)^{p-1}\gamma\left(\sin\cycle +
    2\sin\beta\right) \right]t_2.
\end{align}
The case of $\beta=0$ corresponds to \eq{t-int-bulk}.  The 4D Bloch
Hamiltonian $H_{\rm bulk}(k_x, k_y, k_z, \cycle)$ has four Dirac
points at which the top two valence bands and bottom two conduction
bands all become degenerate at the H~point in the 3D BZ for
$\phi=3\pi/2$. These occur at $\beta=\pm\pi/6$ and $\beta=\pm5\pi/6$,
delineating different regions of $\beta$ with different second Chern
numbers.  In particular, the second Chern number is $\pm1$ when
$\beta\in[-\pi/6,\pi/6]$ or $\beta\in[5\pi/6,7\pi/6]$ respectively,
and is zero otherwise. This is illustrated in Fig.~\ref{fig:dirac},
where the three cases of $\beta\in\{0, \pi/6, \pi/2\}$ are shown.  The
change from a trivial to a nontrivial pumping cycle at $\beta=\pi/6$
is depicted in the two bottom panels. In the second half of the cycle
the bulk HWFs remain localized near the atomic layers, and at
$\phi=3\pi/2$ they exchange Berry flux with the two conduction bands
(rather than with one another) through the Dirac point, changing their
Chern numbers from $\{ +1,-1\}$ to $\{0,0\}$.  In this process, the
net Berry flux through the nonchiral Dirac point vanishes as expected.

Alternatively, one can view the same process in terms of a 5D
Hamiltonian $H_{\rm bulk}(k_x, k_y, k_z, \cycle, \beta)$ where the
second Chern number is calculated on 4D hyperslices corresponding to
fixed values of $\beta$. When $\beta$ passes through the four Dirac
points, the second Chern number of the slices changes by one. The
Dirac points in 5D are the analogs of Weyl points in 3D; indeed, the
integer obtained by integrating the second Chern class on a closed 4D
hypersurface surrounding the degeneracy point is nothing other than
the change in the second Chern number of the hyperslices.

\bibliography{bib}

\begin{thebibliography}{39}%
\makeatletter
\providecommand \@ifxundefined [1]{%
 \@ifx{#1\undefined}
}%
\providecommand \@ifnum [1]{%
 \ifnum #1\expandafter \@firstoftwo
 \else \expandafter \@secondoftwo
 \fi
}%
\providecommand \@ifx [1]{%
 \ifx #1\expandafter \@firstoftwo
 \else \expandafter \@secondoftwo
 \fi
}%
\providecommand \natexlab [1]{#1}%
\providecommand \enquote  [1]{``#1''}%
\providecommand \bibnamefont  [1]{#1}%
\providecommand \bibfnamefont [1]{#1}%
\providecommand \citenamefont [1]{#1}%
\providecommand \href@noop [0]{\@secondoftwo}%
\providecommand \href [0]{\begingroup \@sanitize@url \@href}%
\providecommand \@href[1]{\@@startlink{#1}\@@href}%
\providecommand \@@href[1]{\endgroup#1\@@endlink}%
\providecommand \@sanitize@url [0]{\catcode `\\12\catcode `\$12\catcode
  `\&12\catcode `\#12\catcode `\^12\catcode `\_12\catcode `\%12\relax}%
\providecommand \@@startlink[1]{}%
\providecommand \@@endlink[0]{}%
\providecommand \url  [0]{\begingroup\@sanitize@url \@url }%
\providecommand \@url [1]{\endgroup\@href {#1}{\urlprefix }}%
\providecommand \urlprefix  [0]{URL }%
\providecommand \Eprint [0]{\href }%
\providecommand \doibase [0]{http://dx.doi.org/}%
\providecommand \selectlanguage [0]{\@gobble}%
\providecommand \bibinfo  [0]{\@secondoftwo}%
\providecommand \bibfield  [0]{\@secondoftwo}%
\providecommand \translation [1]{[#1]}%
\providecommand \BibitemOpen [0]{}%
\providecommand \bibitemStop [0]{}%
\providecommand \bibitemNoStop [0]{.\EOS\space}%
\providecommand \EOS [0]{\spacefactor3000\relax}%
\providecommand \BibitemShut  [1]{\csname bibitem#1\endcsname}%
\let\auto@bib@innerbib\@empty
\bibitem [{\citenamefont {Peccei}\ and\ \citenamefont
  {Quinn}(1977)}]{peccei-prl77}%
  \BibitemOpen
  \bibfield  {author} {\bibinfo {author} {\bibfnamefont {R.~D.}\ \bibnamefont
  {Peccei}}\ and\ \bibinfo {author} {\bibfnamefont {H.~R.}\ \bibnamefont
  {Quinn}},\ }\bibfield  {title} {\enquote {\bibinfo {title} {{$\mathrm{CP}$
  Conservation in the Presence of Pseudoparticles}},}\ }\href {\doibase
  10.1103/PhysRevLett.38.1440} {\bibfield  {journal} {\bibinfo  {journal}
  {Phys. Rev. Lett.}\ }\textbf {\bibinfo {volume} {38}},\ \bibinfo {pages}
  {1440} (\bibinfo {year} {1977})}\BibitemShut {NoStop}%
\bibitem [{\citenamefont {Pendlebury}\ \emph {et~al.}(2015)\citenamefont
  {Pendlebury} \emph {et~al.}}]{pendlebury-prd15}%
  \BibitemOpen
  \bibfield  {author} {\bibinfo {author} {\bibfnamefont {J.~M.}\ \bibnamefont
  {Pendlebury}} \emph {et~al.},\ }\bibfield  {title} {\enquote {\bibinfo
  {title} {{Revised experimental upper limit on the electric dipole moment of
  the neutron}},}\ }\href {\doibase 10.1103/PhysRevD.92.092003} {\bibfield
  {journal} {\bibinfo  {journal} {Phys. Rev. D}\ }\textbf {\bibinfo {volume}
  {92}},\ \bibinfo {pages} {092003} (\bibinfo {year} {2015})}\BibitemShut
  {NoStop}%
\bibitem [{\citenamefont {Weinberg}(1978)}]{weinberg-prl78}%
  \BibitemOpen
  \bibfield  {author} {\bibinfo {author} {\bibfnamefont {S.}~\bibnamefont
  {Weinberg}},\ }\bibfield  {title} {\enquote {\bibinfo {title} {{A New Light
  Boson?}}}\ }\href {\doibase 10.1103/PhysRevLett.40.223} {\bibfield  {journal}
  {\bibinfo  {journal} {Phys. Rev. Lett.}\ }\textbf {\bibinfo {volume} {40}},\
  \bibinfo {pages} {223} (\bibinfo {year} {1978})}\BibitemShut {NoStop}%
\bibitem [{\citenamefont {Wilczek}(1978)}]{wilczek-prl78}%
  \BibitemOpen
  \bibfield  {author} {\bibinfo {author} {\bibfnamefont {F.}~\bibnamefont
  {Wilczek}},\ }\bibfield  {title} {\enquote {\bibinfo {title} {{Problem of
  Strong $P$ and $T$ Invariance in the Presence of Instantons}},}\ }\href
  {\doibase 10.1103/PhysRevLett.40.279} {\bibfield  {journal} {\bibinfo
  {journal} {Phys. Rev. Lett.}\ }\textbf {\bibinfo {volume} {40}},\ \bibinfo
  {pages} {279} (\bibinfo {year} {1978})}\BibitemShut {NoStop}%
\bibitem [{\citenamefont {Asztalos}\ \emph {et~al.}(2010)\citenamefont
  {Asztalos}, \citenamefont {Carosi}, \citenamefont {Hagmann}, \citenamefont
  {Kinion}, \citenamefont {van Bibber}, \citenamefont {Hotz}, \citenamefont
  {Rosenberg}, \citenamefont {Rybka}, \citenamefont {Hoskins}, \citenamefont
  {Hwang}, \citenamefont {Sikivie}, \citenamefont {Tanner}, \citenamefont
  {Bradley},\ and\ \citenamefont {Clarke}}]{asztalos-prl10}%
  \BibitemOpen
  \bibfield  {author} {\bibinfo {author} {\bibfnamefont {S.~J.}\ \bibnamefont
  {Asztalos}}, \bibinfo {author} {\bibfnamefont {G.}~\bibnamefont {Carosi}},
  \bibinfo {author} {\bibfnamefont {C.}~\bibnamefont {Hagmann}}, \bibinfo
  {author} {\bibfnamefont {D.}~\bibnamefont {Kinion}}, \bibinfo {author}
  {\bibfnamefont {K.}~\bibnamefont {van Bibber}}, \bibinfo {author}
  {\bibfnamefont {M.}~\bibnamefont {Hotz}}, \bibinfo {author} {\bibfnamefont
  {L.~J}\ \bibnamefont {Rosenberg}}, \bibinfo {author} {\bibfnamefont
  {G.}~\bibnamefont {Rybka}}, \bibinfo {author} {\bibfnamefont
  {J.}~\bibnamefont {Hoskins}}, \bibinfo {author} {\bibfnamefont
  {J.}~\bibnamefont {Hwang}}, \bibinfo {author} {\bibfnamefont
  {P.}~\bibnamefont {Sikivie}}, \bibinfo {author} {\bibfnamefont {D.~B.}\
  \bibnamefont {Tanner}}, \bibinfo {author} {\bibfnamefont {R.}~\bibnamefont
  {Bradley}}, \ and\ \bibinfo {author} {\bibfnamefont {J.}~\bibnamefont
  {Clarke}},\ }\bibfield  {title} {\enquote {\bibinfo {title} {{SQUID-Based
  Microwave Cavity Search for Dark-Matter Axions}},}\ }\href {\doibase
  10.1103/PhysRevLett.104.041301} {\bibfield  {journal} {\bibinfo  {journal}
  {Phys. Rev. Lett.}\ }\textbf {\bibinfo {volume} {104}},\ \bibinfo {pages}
  {041301} (\bibinfo {year} {2010})}\BibitemShut {NoStop}%
\bibitem [{\citenamefont {Wilczek}(1987)}]{wilczek-prl87}%
  \BibitemOpen
  \bibfield  {author} {\bibinfo {author} {\bibfnamefont {F.}~\bibnamefont
  {Wilczek}},\ }\bibfield  {title} {\enquote {\bibinfo {title} {{Two
  applications of axion electrodynamics}},}\ }\href {\doibase
  10.1103/PhysRevLett.58.1799} {\bibfield  {journal} {\bibinfo  {journal}
  {Phys. Rev. Lett.}\ }\textbf {\bibinfo {volume} {58}},\ \bibinfo {pages}
  {1799} (\bibinfo {year} {1987})}\BibitemShut {NoStop}%
\bibitem [{\citenamefont {Witten}(2016)}]{witten-rmp16}%
  \BibitemOpen
  \bibfield  {author} {\bibinfo {author} {\bibfnamefont {E.}~\bibnamefont
  {Witten}},\ }\bibfield  {title} {\enquote {\bibinfo {title} {{Fermion path
  integrals and topological phases}},}\ }\href {\doibase
  10.1103/RevModPhys.88.035001} {\bibfield  {journal} {\bibinfo  {journal}
  {Rev. Mod. Phys.}\ }\textbf {\bibinfo {volume} {88}},\ \bibinfo {pages}
  {035001} (\bibinfo {year} {2016})}\BibitemShut {NoStop}%
\bibitem [{\citenamefont {Gabadadze}\ and\ \citenamefont
  {Shifman}(2002)}]{gabadadze-ijmp87}%
  \BibitemOpen
  \bibfield  {author} {\bibinfo {author} {\bibfnamefont {G.}~\bibnamefont
  {Gabadadze}}\ and\ \bibinfo {author} {\bibfnamefont {M}~\bibnamefont
  {Shifman}},\ }\bibfield  {title} {\enquote {\bibinfo {title} {{QCD vaccum and
  axions: What's happening?}}}\ }\href {\doibase 10.1142/S0217751X02011357}
  {\bibfield  {journal} {\bibinfo  {journal} {Int. J. Mod. Phys. A}\ }\textbf
  {\bibinfo {volume} {17}},\ \bibinfo {pages} {3689} (\bibinfo {year}
  {2002})}\BibitemShut {NoStop}%
\bibitem [{\citenamefont {Wilczek}(2016)}]{wilczek-ps16}%
  \BibitemOpen
  \bibfield  {author} {\bibinfo {author} {\bibfnamefont {F.}~\bibnamefont
  {Wilczek}},\ }\bibfield  {title} {\enquote {\bibinfo {title} {{Particle
  Physics and Condensed Matter: The Saga Continues}},}\ }\href {\doibase
  10.1088/0031-8949/T168/1/014003} {\bibfield  {journal} {\bibinfo  {journal}
  {Phys. Scr.}\ }\textbf {\bibinfo {volume} {T168}},\ \bibinfo {pages} {014003}
  (\bibinfo {year} {2016})}\BibitemShut {NoStop}%
\bibitem [{\citenamefont {Qi}\ \emph {et~al.}(2008)\citenamefont {Qi},
  \citenamefont {Hughes},\ and\ \citenamefont {Zhang}}]{qi2008}%
  \BibitemOpen
  \bibfield  {author} {\bibinfo {author} {\bibfnamefont {X.-L.}\ \bibnamefont
  {Qi}}, \bibinfo {author} {\bibfnamefont {T.~L.}\ \bibnamefont {Hughes}}, \
  and\ \bibinfo {author} {\bibfnamefont {S.-C.}\ \bibnamefont {Zhang}},\
  }\bibfield  {title} {\enquote {\bibinfo {title} {{Topological field theory of
  time-reversal invariant insulators}},}\ }\href {\doibase
  10.1103/PhysRevB.78.195424} {\bibfield  {journal} {\bibinfo  {journal} {Phys.
  Rev. B}\ }\textbf {\bibinfo {volume} {78}},\ \bibinfo {pages} {195424}
  (\bibinfo {year} {2008})}\BibitemShut {NoStop}%
\bibitem [{\citenamefont {Wan}\ \emph {et~al.}(2011)\citenamefont {Wan},
  \citenamefont {Turner}, \citenamefont {Vishwanath},\ and\ \citenamefont
  {Savrasov}}]{wan-prb11}%
  \BibitemOpen
  \bibfield  {author} {\bibinfo {author} {\bibfnamefont {X.}~\bibnamefont
  {Wan}}, \bibinfo {author} {\bibfnamefont {A.~M.}\ \bibnamefont {Turner}},
  \bibinfo {author} {\bibfnamefont {A.}~\bibnamefont {Vishwanath}}, \ and\
  \bibinfo {author} {\bibfnamefont {S.~Y}\ \bibnamefont {Savrasov}},\
  }\bibfield  {title} {\enquote {\bibinfo {title} {{Topological semimetal and
  Fermi-arc surface states in the electronic structure of pyrochlore
  iridates}},}\ }\href {\doibase 10.1103/PhysRevB.83.205101} {\bibfield
  {journal} {\bibinfo  {journal} {Phys. Rev. B}\ }\textbf {\bibinfo {volume}
  {83}},\ \bibinfo {pages} {205101} (\bibinfo {year} {2011})}\BibitemShut
  {NoStop}%
\bibitem [{\citenamefont {Balents}(2011)}]{balents-p11}%
  \BibitemOpen
  \bibfield  {author} {\bibinfo {author} {\bibfnamefont {L.}~\bibnamefont
  {Balents}},\ }\bibfield  {title} {\enquote {\bibinfo {title} {{Viewpoint:
  Weyl electrons kiss}},}\ }\href {http://physics.aps.org/articles/v4/36}
  {\bibfield  {journal} {\bibinfo  {journal} {Physics}\ }\textbf {\bibinfo
  {volume} {4}},\ \bibinfo {pages} {36} (\bibinfo {year} {2011})}\BibitemShut
  {NoStop}%
\bibitem [{\citenamefont {Fang}\ \emph {et~al.}(2012)\citenamefont {Fang},
  \citenamefont {Gilbert},\ and\ \citenamefont {Bernevig}}]{fang-prb12}%
  \BibitemOpen
  \bibfield  {author} {\bibinfo {author} {\bibfnamefont {C.}~\bibnamefont
  {Fang}}, \bibinfo {author} {\bibfnamefont {M.~J.}\ \bibnamefont {Gilbert}}, \
  and\ \bibinfo {author} {\bibfnamefont {B.~A.}\ \bibnamefont {Bernevig}},\
  }\bibfield  {title} {\enquote {\bibinfo {title} {{Bulk topological invariants
  in noninteracting point group symmetric insulators}},}\ }\href {\doibase
  10.1103/PhysRevB.86.115112} {\bibfield  {journal} {\bibinfo  {journal} {Phys.
  Rev. B}\ }\textbf {\bibinfo {volume} {86}},\ \bibinfo {pages} {115112}
  (\bibinfo {year} {2012})}\BibitemShut {NoStop}%
\bibitem [{\citenamefont {Coh}\ \emph {et~al.}(2011)\citenamefont {Coh},
  \citenamefont {Vanderbilt}, \citenamefont {Malashevich},\ and\ \citenamefont
  {Souza}}]{coh-prb11}%
  \BibitemOpen
  \bibfield  {author} {\bibinfo {author} {\bibfnamefont {S.}~\bibnamefont
  {Coh}}, \bibinfo {author} {\bibfnamefont {D.}~\bibnamefont {Vanderbilt}},
  \bibinfo {author} {\bibfnamefont {A.}~\bibnamefont {Malashevich}}, \ and\
  \bibinfo {author} {\bibfnamefont {I.}~\bibnamefont {Souza}},\ }\bibfield
  {title} {\enquote {\bibinfo {title} {{Chern-Simons orbital magnetoelectric
  coupling in generic insulators}},}\ }\href {\doibase
  10.1103/PhysRevB.83.085108} {\bibfield  {journal} {\bibinfo  {journal} {Phys.
  Rev. B}\ }\textbf {\bibinfo {volume} {83}},\ \bibinfo {pages} {085108}
  (\bibinfo {year} {2011})}\BibitemShut {NoStop}%
\bibitem [{\citenamefont {Malashevich}\ \emph {et~al.}(2010)\citenamefont
  {Malashevich}, \citenamefont {Souza}, \citenamefont {Coh},\ and\
  \citenamefont {Vanderbilt}}]{malashevich-njp10}%
  \BibitemOpen
  \bibfield  {author} {\bibinfo {author} {\bibfnamefont {A.}~\bibnamefont
  {Malashevich}}, \bibinfo {author} {\bibfnamefont {I.}~\bibnamefont {Souza}},
  \bibinfo {author} {\bibfnamefont {S.}~\bibnamefont {Coh}}, \ and\ \bibinfo
  {author} {\bibfnamefont {D.}~\bibnamefont {Vanderbilt}},\ }\bibfield  {title}
  {\enquote {\bibinfo {title} {{Theory of orbital magnetoelectric response}},}\
  }\href {\doibase 10.1088/1367-2630/12/5/053032} {\bibfield  {journal}
  {\bibinfo  {journal} {New J. Phys.}\ }\textbf {\bibinfo {volume} {12}},\
  \bibinfo {pages} {053032} (\bibinfo {year} {2010})}\BibitemShut {NoStop}%
\bibitem [{\citenamefont {Essin}\ \emph {et~al.}(2010)\citenamefont {Essin},
  \citenamefont {Turner}, \citenamefont {Moore},\ and\ \citenamefont
  {Vanderbilt}}]{essin2010}%
  \BibitemOpen
  \bibfield  {author} {\bibinfo {author} {\bibfnamefont {A.~M.}\ \bibnamefont
  {Essin}}, \bibinfo {author} {\bibfnamefont {A.~M.}\ \bibnamefont {Turner}},
  \bibinfo {author} {\bibfnamefont {J.~E.}\ \bibnamefont {Moore}}, \ and\
  \bibinfo {author} {\bibfnamefont {D.}~\bibnamefont {Vanderbilt}},\ }\bibfield
   {title} {\enquote {\bibinfo {title} {{Orbital magnetoelectric coupling in
  band insulators}},}\ }\href {\doibase 10.1103/PhysRevB.81.205104} {\bibfield
  {journal} {\bibinfo  {journal} {Phys. Rev. B}\ }\textbf {\bibinfo {volume}
  {81}},\ \bibinfo {pages} {205104} (\bibinfo {year} {2010})}\BibitemShut
  {NoStop}%
\bibitem [{\citenamefont {Essin}\ \emph {et~al.}(2009)\citenamefont {Essin},
  \citenamefont {Moore},\ and\ \citenamefont {Vanderbilt}}]{essin-prl09}%
  \BibitemOpen
  \bibfield  {author} {\bibinfo {author} {\bibfnamefont {A.~M.}\ \bibnamefont
  {Essin}}, \bibinfo {author} {\bibfnamefont {J.~E.}\ \bibnamefont {Moore}}, \
  and\ \bibinfo {author} {\bibfnamefont {D.}~\bibnamefont {Vanderbilt}},\
  }\bibfield  {title} {\enquote {\bibinfo {title} {{Magnetoelectric
  Polarizability and Axion Electrodynamics in Crystalline Insulators}},}\
  }\href {\doibase 10.1103/PhysRevLett.102.146805} {\bibfield  {journal}
  {\bibinfo  {journal} {Phys. Rev. Lett.}\ }\textbf {\bibinfo {volume} {102}},\
  \bibinfo {pages} {146805} (\bibinfo {year} {2009})}\BibitemShut {NoStop}%
\bibitem [{\citenamefont {King-Smith}\ and\ \citenamefont
  {Vanderbilt}(1993)}]{king-smith-prb93}%
  \BibitemOpen
  \bibfield  {author} {\bibinfo {author} {\bibfnamefont {R.~D.}\ \bibnamefont
  {King-Smith}}\ and\ \bibinfo {author} {\bibfnamefont {D.}~\bibnamefont
  {Vanderbilt}},\ }\bibfield  {title} {\enquote {\bibinfo {title} {{Theory of
  polarization of crystalline solids}},}\ }\href {\doibase
  10.1103/PhysRevB.47.1651} {\bibfield  {journal} {\bibinfo  {journal} {Phys.
  Rev. B}\ }\textbf {\bibinfo {volume} {47}},\ \bibinfo {pages} {1651}
  (\bibinfo {year} {1993})}\BibitemShut {NoStop}%
\bibitem [{\citenamefont {Vanderbilt}\ and\ \citenamefont
  {King-Smith}(1993)}]{vanderbilt-prb93}%
  \BibitemOpen
  \bibfield  {author} {\bibinfo {author} {\bibfnamefont {D.}~\bibnamefont
  {Vanderbilt}}\ and\ \bibinfo {author} {\bibfnamefont {R.~D.}\ \bibnamefont
  {King-Smith}},\ }\bibfield  {title} {\enquote {\bibinfo {title} {{Electric
  polarization as a bulk quantity and its relation to surface charge}},}\
  }\href {\doibase 10.1103/PhysRevB.48.4442} {\bibfield  {journal} {\bibinfo
  {journal} {Phys. Rev. B}\ }\textbf {\bibinfo {volume} {48}},\ \bibinfo
  {pages} {4442} (\bibinfo {year} {1993})}\BibitemShut {NoStop}%
\bibitem [{\citenamefont {Taherinejad}\ and\ \citenamefont
  {Vanderbilt}(2015)}]{taherinejad-prl15}%
  \BibitemOpen
  \bibfield  {author} {\bibinfo {author} {\bibfnamefont {M.}~\bibnamefont
  {Taherinejad}}\ and\ \bibinfo {author} {\bibfnamefont {D.}~\bibnamefont
  {Vanderbilt}},\ }\bibfield  {title} {\enquote {\bibinfo {title} {{Adiabatic
  Pumping of Chern-Simons Axion Coupling}},}\ }\href {\doibase
  10.1103/PhysRevLett.114.096401} {\bibfield  {journal} {\bibinfo  {journal}
  {Phys. Rev. Lett.}\ }\textbf {\bibinfo {volume} {114}},\ \bibinfo {pages}
  {096401} (\bibinfo {year} {2015})}\BibitemShut {NoStop}%
\bibitem [{\citenamefont {Marzari}\ and\ \citenamefont
  {Vanderbilt}(1997)}]{marzari-prb97}%
  \BibitemOpen
  \bibfield  {author} {\bibinfo {author} {\bibfnamefont {N.}~\bibnamefont
  {Marzari}}\ and\ \bibinfo {author} {\bibfnamefont {D.}~\bibnamefont
  {Vanderbilt}},\ }\bibfield  {title} {\enquote {\bibinfo {title} {{Maximally
  localized generalized Wannier functions for composite energy bands}},}\
  }\href {\doibase 10.1103/PhysRevB.56.12847} {\bibfield  {journal} {\bibinfo
  {journal} {Phys. Rev. B}\ }\textbf {\bibinfo {volume} {56}},\ \bibinfo
  {pages} {12847} (\bibinfo {year} {1997})}\BibitemShut {NoStop}%
\bibitem [{\citenamefont {Taherinejad}\ \emph {et~al.}(2014)\citenamefont
  {Taherinejad}, \citenamefont {Garrity},\ and\ \citenamefont
  {Vanderbilt}}]{taherinejad-prb14}%
  \BibitemOpen
  \bibfield  {author} {\bibinfo {author} {\bibfnamefont {M.}~\bibnamefont
  {Taherinejad}}, \bibinfo {author} {\bibfnamefont {K.~F.}\ \bibnamefont
  {Garrity}}, \ and\ \bibinfo {author} {\bibfnamefont {D.}~\bibnamefont
  {Vanderbilt}},\ }\bibfield  {title} {\enquote {\bibinfo {title} {{Wannier
  center sheets in topological insulators}},}\ }\href {\doibase
  10.1103/PhysRevB.89.115102} {\bibfield  {journal} {\bibinfo  {journal} {Phys.
  Rev. B}\ }\textbf {\bibinfo {volume} {89}},\ \bibinfo {pages} {115102}
  (\bibinfo {year} {2014})}\BibitemShut {NoStop}%
\bibitem [{Note1()}]{Note1}%
  \BibitemOpen
  \bibinfo {note} {The expression for $\theta _{\Delta xy}$ in Eq.~(12) of
  Ref.~\protect \rev@citealpnum {taherinejad-prl15} has a typo in it: a minus
  sign is missing.}\BibitemShut {Stop}%
\bibitem [{\citenamefont {Ceresoli}\ \emph {et~al.}(2006)\citenamefont
  {Ceresoli}, \citenamefont {Thonhauser}, \citenamefont {Vanderbilt},\ and\
  \citenamefont {Resta}}]{ceresoli-prb06}%
  \BibitemOpen
  \bibfield  {author} {\bibinfo {author} {\bibfnamefont {D.}~\bibnamefont
  {Ceresoli}}, \bibinfo {author} {\bibfnamefont {T.}~\bibnamefont
  {Thonhauser}}, \bibinfo {author} {\bibfnamefont {D.}~\bibnamefont
  {Vanderbilt}}, \ and\ \bibinfo {author} {\bibfnamefont {R.}~\bibnamefont
  {Resta}},\ }\bibfield  {title} {\enquote {\bibinfo {title} {{Orbital
  magnetization in crystalline solids: Multi-band insulators, Chern insulators,
  and metals}},}\ }\href {\doibase 10.1103/PhysRevB.74.024408} {\bibfield
  {journal} {\bibinfo  {journal} {Phys. Rev. B}\ }\textbf {\bibinfo {volume}
  {74}},\ \bibinfo {pages} {024408} (\bibinfo {year} {2006})}\BibitemShut
  {NoStop}%
\bibitem [{\citenamefont {Blount}(1962)}]{blount-ssp62}%
  \BibitemOpen
  \bibfield  {author} {\bibinfo {author} {\bibfnamefont {E.~I.}\ \bibnamefont
  {Blount}},\ }\bibfield  {title} {\enquote {\bibinfo {title} {{Formalisms of
  Band Theory}},}\ }\href@noop {} {\bibfield  {journal} {\bibinfo  {journal}
  {Solid State Phys.}\ }\textbf {\bibinfo {volume} {13}},\ \bibinfo {pages}
  {305} (\bibinfo {year} {1962})}\BibitemShut {NoStop}%
\bibitem [{\citenamefont {Mead}(1992)}]{mead-rmp92}%
  \BibitemOpen
  \bibfield  {author} {\bibinfo {author} {\bibfnamefont {C.~A.}\ \bibnamefont
  {Mead}},\ }\bibfield  {title} {\enquote {\bibinfo {title} {{The geometric
  phase in molecular systems}},}\ }\href {\doibase 10.1103/RevModPhys.64.51}
  {\bibfield  {journal} {\bibinfo  {journal} {Rev. Mod. Phys.}\ }\textbf
  {\bibinfo {volume} {64}},\ \bibinfo {pages} {51} (\bibinfo {year}
  {1992})}\BibitemShut {NoStop}%
\bibitem [{\citenamefont {Bianco}\ and\ \citenamefont
  {Resta}(2011)}]{bianco-prb11}%
  \BibitemOpen
  \bibfield  {author} {\bibinfo {author} {\bibfnamefont {R.}~\bibnamefont
  {Bianco}}\ and\ \bibinfo {author} {\bibfnamefont {R.}~\bibnamefont {Resta}},\
  }\bibfield  {title} {\enquote {\bibinfo {title} {{Mapping topological order
  in coordinate space}},}\ }\href {\doibase 10.1103/PhysRevB.84.241106}
  {\bibfield  {journal} {\bibinfo  {journal} {Phys. Rev. B}\ }\textbf {\bibinfo
  {volume} {84}},\ \bibinfo {pages} {241106} (\bibinfo {year}
  {2011})}\BibitemShut {NoStop}%
\bibitem [{\citenamefont {Haldane}(1988)}]{haldane-prl88}%
  \BibitemOpen
  \bibfield  {author} {\bibinfo {author} {\bibfnamefont {F.~D.~M.}\
  \bibnamefont {Haldane}},\ }\bibfield  {title} {\enquote {\bibinfo {title}
  {{Model for a quantum Hall effect without Landau levels: Condensed-matter
  realization of the ``parity anomaly''}},}\ }\href {\doibase
  10.1103/PhysRevLett.61.2015} {\bibfield  {journal} {\bibinfo  {journal}
  {Phys. Rev. Lett.}\ }\textbf {\bibinfo {volume} {61}},\ \bibinfo {pages}
  {2015} (\bibinfo {year} {1988})}\BibitemShut {NoStop}%
\bibitem [{Note2()}]{Note2}%
  \BibitemOpen
  \bibinfo {note} {Available at \protect \url
  {http://physics.rutgers.edu/pythtb/}.}\BibitemShut {Stop}%
\bibitem [{\citenamefont {Thonhauser}\ and\ \citenamefont
  {Vanderbilt}(2006)}]{thonhauser-prb06}%
  \BibitemOpen
  \bibfield  {author} {\bibinfo {author} {\bibfnamefont {T.}~\bibnamefont
  {Thonhauser}}\ and\ \bibinfo {author} {\bibfnamefont {D.}~\bibnamefont
  {Vanderbilt}},\ }\bibfield  {title} {\enquote {\bibinfo {title}
  {{Insulator/Chern-insulator transition in the Haldane model}},}\ }\href
  {\doibase 10.1103/PhysRevB.74.235111} {\bibfield  {journal} {\bibinfo
  {journal} {Phys. Rev. B}\ }\textbf {\bibinfo {volume} {74}},\ \bibinfo
  {pages} {235111} (\bibinfo {year} {2006})}\BibitemShut {NoStop}%
\bibitem [{Note3()}]{Note3}%
  \BibitemOpen
  \bibinfo {note} {In addition to mirror symmetry $M_z$, the bulk model at
  $\phi =0,\pi $ also exhibits $C_2^y*{\protect \cal T}$ symmetry, with the
  two-fold axis lying on the plane halfway between two atomic layers and
  pointing along a zig-zag edge. This additional symmetry, which also
  constrains $\theta _3$ to be an integer multiple of $\pi $, can be removed by
  changing the neareast-neighbor hopping amplitude along one of the two
  in-plane directions. We have checked that doing so modifies the $\theta
  _3(\phi )$ curve, but does not affect the quantized values at $\phi =0,\pi
  $.}\BibitemShut {Stop}%
\bibitem [{\citenamefont {Varjas}\ \emph {et~al.}(2015)\citenamefont {Varjas},
  \citenamefont {de~Juan},\ and\ \citenamefont {Lu}}]{varjas-prb15}%
  \BibitemOpen
  \bibfield  {author} {\bibinfo {author} {\bibfnamefont {D.}~\bibnamefont
  {Varjas}}, \bibinfo {author} {\bibfnamefont {F.}~\bibnamefont {de~Juan}}, \
  and\ \bibinfo {author} {\bibfnamefont {Y.-M.}\ \bibnamefont {Lu}},\
  }\bibfield  {title} {\enquote {\bibinfo {title} {Bulk invariants and
  topological response in insulators and superconductors with nonsymmorphic
  symmetries},}\ }\href {\doibase 10.1103/PhysRevB.92.195116} {\bibfield
  {journal} {\bibinfo  {journal} {Phys. Rev. B}\ }\textbf {\bibinfo {volume}
  {92}},\ \bibinfo {pages} {195116} (\bibinfo {year} {2015})}\BibitemShut
  {NoStop}%
\bibitem [{\citenamefont {Thouless}(1983)}]{thouless-prb83}%
  \BibitemOpen
  \bibfield  {author} {\bibinfo {author} {\bibfnamefont {D.~J.}\ \bibnamefont
  {Thouless}},\ }\bibfield  {title} {\enquote {\bibinfo {title} {{Quantization
  of particle transport}},}\ }\href {\doibase 10.1103/PhysRevB.27.6083}
  {\bibfield  {journal} {\bibinfo  {journal} {Phys. Rev. B}\ }\textbf {\bibinfo
  {volume} {27}},\ \bibinfo {pages} {6083} (\bibinfo {year}
  {1983})}\BibitemShut {NoStop}%
\bibitem [{\citenamefont {Gos\'albez-Mart\'{\i}nez}\ \emph
  {et~al.}(2015)\citenamefont {Gos\'albez-Mart\'{\i}nez}, \citenamefont
  {Souza},\ and\ \citenamefont {Vanderbilt}}]{gosalbez-prb15}%
  \BibitemOpen
  \bibfield  {author} {\bibinfo {author} {\bibfnamefont {D.}~\bibnamefont
  {Gos\'albez-Mart\'{\i}nez}}, \bibinfo {author} {\bibfnamefont
  {I.}~\bibnamefont {Souza}}, \ and\ \bibinfo {author} {\bibfnamefont
  {D.}~\bibnamefont {Vanderbilt}},\ }\bibfield  {title} {\enquote {\bibinfo
  {title} {{Chiral degeneracies and Fermi-surface Chern numbers in bcc Fe}},}\
  }\href {\doibase 10.1103/PhysRevB.78.045426} {\bibfield  {journal} {\bibinfo
  {journal} {Phys. Rev. B}\ }\textbf {\bibinfo {volume} {92}},\ \bibinfo
  {pages} {085138} (\bibinfo {year} {2015})}\BibitemShut {NoStop}%
\bibitem [{\citenamefont {Fu}\ and\ \citenamefont {Kane}(2006)}]{fu-prb06}%
  \BibitemOpen
  \bibfield  {author} {\bibinfo {author} {\bibfnamefont {L.}~\bibnamefont
  {Fu}}\ and\ \bibinfo {author} {\bibfnamefont {C.~L.}\ \bibnamefont {Kane}},\
  }\bibfield  {title} {\enquote {\bibinfo {title} {{Time reversal polarization
  and a ${Z}_{2}$ adiabatic spin pump}},}\ }\href {\doibase
  10.1103/PhysRevB.74.195312} {\bibfield  {journal} {\bibinfo  {journal} {Phys.
  Rev. B}\ }\textbf {\bibinfo {volume} {74}},\ \bibinfo {pages} {195312}
  (\bibinfo {year} {2006})}\BibitemShut {NoStop}%
\bibitem [{\citenamefont {Soluyanov}\ and\ \citenamefont
  {Vanderbilt}(2011)}]{soluyanov-prb11}%
  \BibitemOpen
  \bibfield  {author} {\bibinfo {author} {\bibfnamefont {A.~A.}\ \bibnamefont
  {Soluyanov}}\ and\ \bibinfo {author} {\bibfnamefont {D.}~\bibnamefont
  {Vanderbilt}},\ }\bibfield  {title} {\enquote {\bibinfo {title} {{Computing
  topological invariants without inversion symmetry}},}\ }\href {\doibase
  10.1103/PhysRevB.83.235401} {\bibfield  {journal} {\bibinfo  {journal} {Phys.
  Rev. B}\ }\textbf {\bibinfo {volume} {83}},\ \bibinfo {pages} {235401}
  (\bibinfo {year} {2011})}\BibitemShut {NoStop}%
\bibitem [{\citenamefont {Yu}\ \emph {et~al.}(2011)\citenamefont {Yu},
  \citenamefont {Qi}, \citenamefont {Bernevig}, \citenamefont {Fang},\ and\
  \citenamefont {Dai}}]{yu-prb11}%
  \BibitemOpen
  \bibfield  {author} {\bibinfo {author} {\bibfnamefont {R.}~\bibnamefont
  {Yu}}, \bibinfo {author} {\bibfnamefont {X.~L.}\ \bibnamefont {Qi}}, \bibinfo
  {author} {\bibfnamefont {A.}~\bibnamefont {Bernevig}}, \bibinfo {author}
  {\bibfnamefont {Z.}~\bibnamefont {Fang}}, \ and\ \bibinfo {author}
  {\bibfnamefont {X.}~\bibnamefont {Dai}},\ }\bibfield  {title} {\enquote
  {\bibinfo {title} {{Equivalent expression of ${\mathbb{Z}}_{2}$ topological
  invariant for band insulators using the non-Abelian Berry connection}},}\
  }\href {\doibase 10.1103/PhysRevB.84.075119} {\bibfield  {journal} {\bibinfo
  {journal} {Phys. Rev. B}\ }\textbf {\bibinfo {volume} {84}},\ \bibinfo
  {pages} {075119} (\bibinfo {year} {2011})}\BibitemShut {NoStop}%
\bibitem [{\citenamefont {Gresch}\ \emph {et~al.}(2016)\citenamefont {Gresch},
  \citenamefont {Aut{\`e}s}, \citenamefont {Yazyev}, \citenamefont {Troyer},
  \citenamefont {Vanderbilt}, \citenamefont {Bernevig},\ and\ \citenamefont
  {Soluyanov}}]{gresch-arxiv16}%
  \BibitemOpen
  \bibfield  {author} {\bibinfo {author} {\bibfnamefont {D.}~\bibnamefont
  {Gresch}}, \bibinfo {author} {\bibfnamefont {G.}~\bibnamefont {Aut{\`e}s}},
  \bibinfo {author} {\bibfnamefont {O.~V.}\ \bibnamefont {Yazyev}}, \bibinfo
  {author} {\bibfnamefont {M.}~\bibnamefont {Troyer}}, \bibinfo {author}
  {\bibfnamefont {D.}~\bibnamefont {Vanderbilt}}, \bibinfo {author}
  {\bibfnamefont {B.~A.}\ \bibnamefont {Bernevig}}, \ and\ \bibinfo {author}
  {\bibfnamefont {A.~A.}\ \bibnamefont {Soluyanov}},\ }\bibfield  {title}
  {\enquote {\bibinfo {title} {{Z2Pack: Numerical Implementation of Hybrid
  Wannier Centers for Identifying Topological Materials}},}\ }\href
  {https://arxiv.org/abs/1610.08983} {\bibfield  {journal} {\bibinfo  {journal}
  {arXiv:1610.08983}\ } (\bibinfo {year} {2016})}\BibitemShut {NoStop}%
\bibitem [{\citenamefont {Teo}\ \emph {et~al.}(2008)\citenamefont {Teo},
  \citenamefont {Fu},\ and\ \citenamefont {Kane}}]{teo-prb08}%
  \BibitemOpen
  \bibfield  {author} {\bibinfo {author} {\bibfnamefont {J.~C.~Y.}\
  \bibnamefont {Teo}}, \bibinfo {author} {\bibfnamefont {Liang}\ \bibnamefont
  {Fu}}, \ and\ \bibinfo {author} {\bibfnamefont {C.~L.}\ \bibnamefont
  {Kane}},\ }\bibfield  {title} {\enquote {\bibinfo {title} {{Surface states
  and topological invariants in three-dimensional topological insulators:
  Application to ${\text{Bi}}_{1\ensuremath{-}x}{\text{Sb}}_{x}$}},}\ }\href
  {\doibase 10.1103/PhysRevB.78.045426} {\bibfield  {journal} {\bibinfo
  {journal} {Phys. Rev. B}\ }\textbf {\bibinfo {volume} {78}},\ \bibinfo
  {pages} {045426} (\bibinfo {year} {2008})}\BibitemShut {NoStop}%
\end{thebibliography}%

\end{document}